\documentclass[11pt,twoside]{article}
\usepackage{atmp04,amssymb,amsmath}

\newcommand{\Tr}{\mathrm{Tr}}
\newcommand{\tr}{\mathrm{tr}}
\newcommand{\A}{\mathcal{A}}

\def\tilde{\widetilde}
\def\bar{\overline}
\def\Z{{\bf Z}}
\def\S{{\bf S}}

\font\zfont = cmss10 
\font\litfont = cmr6

\def\bigone{\hbox{1\kern -.23em {\rm l}}}
\def\ZZ{\hbox{\zfont Z\kern-.4emZ}}
\def\half{{\litfont {\frac{1}{2}}}}

\begin{document}

\copyrightnotice{2003}{7}{577}{617}
\setcounter{page}{577}
\numberwithin{equation}{section}
\pagestyle{myheadings}
\markboth{\it Unification Scale, Proton Decay,$\cdots$}{\it T. Friedmann and E. Witten}
\title{Unification Scale, Proton Decay, And Manifolds Of $G_2$ Holonomy }
\author{Tamar Friedmann$^a$ and Edward Witten$^b$} 
\address{$^a$Princeton University, Princeton NJ 08540 USA\\
 $^b$Institute For Advanced Study, Princeton NJ 08540 USA}
\addressemail{tamarf@mit.edu, witten@ias.edu}
\arxurl{hep-th/0211269}

\begin{abstract} 
Models of particle physics based on manifolds of $G_2$
holonomy are in most respects much more complicated than other
string-derived models, but as we show here they do have one
simplification: threshold corrections to grand unification are
particularly simple.  We compute these corrections, getting
completely explicit results in some simple cases.  We estimate
the relation between Newton's constant, the GUT scale, and the
value of $\alpha_{GUT}$, and explore the implications for proton
decay.  In the case of proton decay, there is an interesting
mechanism which (relative to four-dimensional SUSY GUT's)
enhances the gauge boson contribution to $p\to\pi^0e^+_L$ compared
to other modes such as $p\to \pi^0e^+_R$ or $p\to \pi^+\bar\nu_R$.
Because of numerical uncertainties, we do not know whether to
intepret this as an enhancement of the $p\to \pi^0e^+_L$ mode or
a suppression of the others.
\end{abstract}
\section{Introduction}
\label{sec:1}

In the original estimation of the running of coupling constants in
grand unified theories \cite{gqw}, it was found that the low
energy $SU(3)\times SU(2)\times U(1)$ couplings can be unified in
a simple gauge group such as $SU(5)$ \cite{gg} at an energy of about $10^{15}$ GeV.
This energy is suggestively close to the Planck mass
$M_{Pl}=1.2\times 10^{19}$ GeV, at least on a log scale. This
hints at a unification that includes gravity as well as the other
forces.

Subsequently, with more precise measurements of the low energy
couplings, it became clear that coupling unification occurs much
more precisely in {\it supersymmetric} grand unification.
Supersymmetry raises the unification scale $M_{GUT}$ to about
$2-3\times 10^{16}$ GeV  \cite{rdw}, assuming that the running of
couplings can be computed using only the known particles plus
Higgs bosons and superpartners. Incorporating supersymmetry
reduces the discrepancy between the unification scale and the
Planck scale, measured logarithmically, by about one third.

In this discussion, it is not clear if the unification mass should
be compared precisely to $M_{Pl}$ or, say, to $M_{Pl}/2\pi$. For
this question to make sense, one needs a more precise unified
theory including gravity. The first sufficiently precise model was
the perturbative $E_8\times E_8$ heterotic string \cite{ghmr}. In this model,
the discrepancy between the scales of grand unification and
gravity was again reduced, relative to the original estimates,
essentially because the string scale is somewhat below the Planck
scale.  The remaining discrepancy, evaluated at tree level, is
about a factor of 20 -- that is, the GUT or grand unification
scale as inferred from low energy couplings is about 20 times
smaller than one would expect based on the tree level of the
heterotic string.  For a discussion, see \cite{ginsparg}.

%
 This discrepancy, which is about six percent on a log scale (and
thus roughly a third as large as the mismatch originally estimated
in \cite{gqw}), is small enough to raise the question of how much the
one-loop threshold corrections might close the gap. The threshold
corrections have been calculated {\cite{kaplunovsky}}, generalizing
the corresponding computations in field theory 
{\cite{weinberg}, \cite{OvrutBK}}. 
The threshold corrections depend on the detailed
choice of a heterotic string compactification.  In most simple
models, they seem too small to give the desired effect.

An alternative is to assume additional charged particles with
masses far below the unification scale, so as to modify the
renormalization group estimate of the unification scale. To
preserve the usual SUSY GUT prediction for the weak mixing angle
$\sin^2\theta_W$, one might limit oneself to complete $SU(5)$
multiplets; for assessment of possibilities in this framework, see
{\cite{babupati},\cite{bmpz},\cite{GhilenceaMU}}.

Another alternative is to consider the {\it strongly coupled}
$E_8\times E_8$ heterotic string, in which an eleventh dimension
opens up and gauge fields propagate on the boundaries \cite{hw}. In this case, rather than a prediction for the
relation between Newton's constant $G_N=1/M_{Pl}^2$ and the scale
of grand unification, one gets only an inequality \cite{witten}. The
inequality is difficult to evaluate precisely; the estimate given
in \cite{witten}\ is 
\begin{equation}\label{estom}
{ G_N\geq \frac{\alpha_{GUT}^2}{
16\pi^2}\left|\int_Z\omega\wedge\frac{\tr F\wedge F-\frac{1}{ 2}R\wedge
R }{ 8\pi^2}\right|, }\end{equation}
 where $\omega$ is the Kahler form of the
Calabi-Yau manifold $Z$, and one would expect the integral in
(\ref{estom})\ to be of order $1/M_{GUT}^2$ times a number fairly close
to 1; a further estimate was given in 
{\cite{CurioDW}} , where this bound was lowered by a factor of 2/3. 
 The observed value of
$G_N$ is rather close to saturating this inequality and probably
does so within the uncertainties.\footnote{There is also a special
case, with equal instanton numbers in the two boundaries, in
which one does not obtain such an inequality, and $G_N$ can be
much smaller. This case was not pursued seriously in \cite{witten},
because of a preference to maintain the successes of grand
unification. } If this inequality, or the one derived in the
present paper, is saturated, this is a very interesting statement
about nature, but a theoretical reason to expect the inequality
to be saturated is not clear.

The purpose of the present paper is to study this and related
questions in the context of another related type of model, namely
$M$-theory compactification on a (singular) manifold of $G_2$
holonomy.  Such models can be dual to heterotic strings; in many
(but presumably not all) instances, a compactification of
$M$-theory on a $G_2$ manifold $X$ is dual to the compactification
of a heterotic string on a Calabi-Yau threefold $Z$. So many
models actually have (at least) three different dual descriptions:
as a perturbative $E_8\times E_8$ heterotic string; as a strongly
coupled heterotic string with gauge fields on the boundary; and in
terms of $M$-theory on a singular $G_2$ manifold.
 As in most examples of
duality, when one of these descriptions is useful, the others are
strongly coupled and difficult to use.

Apart from duality with the heterotic string, $G_2$ models that
might be relevant to phenomenology can be constructed via duality
with certain Type II orientifolds %
%
%
 %
 {\cite{cvetic}-\cite{blum}}, which in turn may have
descriptions via the Type I string or $SO(32)$ heterotic string.

A particular approach to constructing  semirealistic particle
physics models based on $G_2$ manifolds was described in
\cite{gwitten}. The ability to 
generate chiral fermions depended on somewhat subtle singularities
that have been
studied in  %
%
%
%
{\cite{pwitten} - \cite{gb}}.  We will pursue this approach further in
the present paper.

$G_2$ manifolds are much more difficult and less fully understood
than the Calabi-Yau threefolds that  can be used in constructing
heterotic string compactifications. However, as we will see, they
have at least one nice simplification relative to the generic
Calabi-Yau models: the threshold corrections are rather simple in
the case of compactification on a $G_2$ manifold.  They are in
fact given by a topological invariant, the Ray-Singer analytic
torsion
%
%
{\cite{ray},\cite{rs}}.  As a result, the threshold corrections can be
computed explicitly based on topological assumptions, without
needing to know the detailed form of the $G_2$ metric.  That is
convenient, since explicit $G_2$ metrics are unknown.

In this paper, we will carry out for $G_2$ manifolds as much as we
can of the usual program of grand unification.  We compute the
threshold corrections, obtain as precise an estimate as we can for
the radius of compactification, derive the inequality which is
analogous to (\ref{estom}), and attempt to estimate the proton lifetime.
Of course, there is some serious model dependence here; we attempt
to make statements that apply to classes of interesting models.
For example, to discuss proton decay in a sensible fashion, we
need to assume a mechanism for doublet-triplet splitting; here, we
assume the mechanism of \cite{gwitten}\ (which in turn is a close cousin
of the original mechanism for doublet-triplet splitting in
Calabi-Yau compactification of the heterotic string \cite{chsw}; this mechanism has been
reconsidered recently from a bottom-up point of view
%
%
%
%
%
{\cite{bottomup} - \cite{hmn}}). One consequence of the particular
doublet-triplet splitting mechanism that we use is that dimension
five (or four) operators contributing to proton decay are absent;
proton decay is dominated, therefore, by dimension six operators.
In four dimensions, those operators are induced by gauge boson and
Higgs boson exchange; we will see that in the present context, in
a sense, there is a direct $M$-theory contribution.  We will focus
in this paper on $SU(5)$ models, partly for simplicity and partly
because the doublet-triplet splitting mechanism of \cite{gwitten}\ has
been formulated for this case.

The results we get are as follows:

(1)  For $SU(5)$ models, the threshold corrections due to
Kaluza-Klein harmonics do not modify the prediction for
$\sin^2\theta_W$ obtained from four-dimensional supersymmetric
GUT's.  (This is somewhat analogous to a similar result for
massive gauge bosons in four-dimensional $SU(5)$ models.) But they
do modify the unification scale; that is, they modify the relation
between the compactification scale and the scale of grand
unification as estimated from low energy gauge couplings. This
modification is significant in estimating the predictions for
Newton's constant and for proton decay.

(2)  The inequality analogous to (\ref{estom})\ is again rather close to
being saturated and perhaps even more so.  The details depend on
the threshold corrections as well as on other factors that will
appear in the discussion.

(3)  In four-dimensional SUSY GUT's, the gauge boson exchange
contributions to proton decay, if they dominate, lead to a proton
lifetime that has been estimated recently as $5\times 10^{36\pm 1}$ years
\cite{raby}. (Four-dimensional GUT's also generally have a
dimension five contribution that can be phenomenologically
troublesome;
%
 it is absent in the models we consider. See also {\cite{BajcBV}} for another approach.)  Relative to
four-dimensional GUT's, we find a mechanism that enhances proton
decay modes such as\footnote{And similar modes with $\pi^0$ replaced
by $K^0$, or $e^+_L$ by $\mu^+_L$.  As in most four-dimensional
GUT's, we do not have precise knowledge of the flavor structure.}
$p\to \pi^0e^+_L$ relative to $p\to \pi^0e^+_R$ or $p\to
\pi^+\bar\nu_R$; these are typical modes that arise from gauge
boson exchange. Given numerical uncertainties that will appear in
section 5, it is hard to say if in practice this mechanism
enhances the $p\to \pi^0e^+_L$ modes relative to GUT's or
suppresses the others. Because of these issues, to cite a very
rough estimate of the proton lifetime, we somewhat arbitrarily
keep the central value estimated in the four-dimensional models
and double the logarithmic uncertainty, so that the proton
lifetime in these models if gauge boson exchange dominates might
be $5\times 10^{36\pm 2}$ years.\footnote{The current experimental bound on
$p\to \pi^0 e^+$ is $4.4 \times 10^{33}$ years (for a recent
report, see \cite{ganezer}) and the next generation of experiments may improve this by
a factor of 10 to 20 (for example, see \cite{jung}).} (In practice, the uncertainty
could be much larger if because of additional light charged
particles, the unification scale is modified; the same statement
applies to four-dimensional GUT models.)

This paper is organized as follows.  In section 2, we review how
semi-realistic models of particle physics can be obtained by
compactification of $M$-theory on a singular manifold of $G_2$
holonomy. In section 3, we express the threshold corrections in
terms of Ray-Singer torsion and make everything completely
explicit in the case of  a lens space.   In section 4, we work out
the implications for Newton's constant. In section 5, we discuss
proton decay.  In section 6, we discuss some unresolved issues
about these models.
 Finally, in the appendices, we do some calculations of torsion to
 fill in some loose ends in section 3.

\section{Review Of Models}
\label{sec:2}

As we will now recall, duality with the heterotic string implies
that semi-realistic models of particle physics can be derived by
compactification of $M$-theory on a manifold of $G_2$ holonomy.
(One could also use Type II orientifolds as the starting point
here {\cite{cvetic}-\cite{blum}}.) Semi-realistic means that gauge groups
and fermion quantum numbers come out correctly, but there is no
good understanding of things that depend on supersymmetry
breaking, like the correct fixing of moduli, as well as  fermion
masses (which depend on the moduli).

In fact, semi-realistic  models of particle physics can be derived
by compactification of the $E_8\times E_8$ heterotic string on a
Calabi-Yau threefold $Z$.  Most Calabi-Yau threefolds participate
in mirror symmetry.  Mirror symmetry is interpreted to mean
\cite{syz} that in a suitable region of its moduli space,
$Z$ is fibered over a three-manifold $Q$ with fibers that are
generically copies of a three-torus ${\bf T}^3$.

Now, we take $Q$ to be very large in string units, while keeping
the size of the ${\bf T}^3$ fixed and taking the string coupling
constant to infinity.  The strong coupling limit of the heterotic
string on ${\bf T}^3$ is $M$-theory on K3.  So in this limit, we
replace all of the ${\bf T}^3$'s by K3's.  The heterotic string
on $Z$ then turns into $M$-theory on a seven-manifold $X$ that is
fibered over $Q$ with generic fibers being copies of K3.
Supersymmetry requires that $X$ actually has $G_2$ holonomy.

Suppose that the original heterotic string model had an unbroken
gauge group $G\subset E_8\times E_8$.  If $G$ is simple, for
example  $G=SU(5)$, then it is a group of type A, D, or E (as
these are the simple gauge groups that occur in heterotic string
models at level one), and will appear in $M$-theory on K3 as an
orbifold singularity of the appropriate type. Such a singularity
will appear in each fiber of the fibration $X\to Q$. They will fit
together into a copy of $Q$ embedded in $X$ along which $X$ has
the appropriate A, D, or E orbifold singularity. In the present
paper, when we want to be specific, we consider $G=SU(5)$, in
which case $Q$ is a locus of $\Z_5$ orbifold singularities.  This
means, concretely, that the normal space to $Q$ in $X$ can be
parametrized locally by complex coordinates $w_1,$ $w_2$ with the
identification 
\begin{equation}\label{loppy}
{(w_1,w_2)\to (e^{2\pi i/5}w_1,e^{-2\pi
i/5}w_2).}\end{equation}

Though our focus in this paper is on the case that $G$ is simple,
we briefly note that if $G$ is semi-simple, this means that the
K3 contains several disjoint singularities, one for each factor
in $G$.  These will lead, generally, to different components of
the singular set of $X$.  (They might intersect, in which case
chiral superfields may be supported at the points of
intersection.) Also, $U(1)$ factors in $G$ appear in $M$-theory
as modes of the three-form field $C$ of $M$-theory.

We can abstract what we have learned from the way that we have
learned it and say that in $M$-theory on a $G_2$ manifold $X$,
gauge theory of type A, D, or E arises from the existence on $X$
of a three-dimensional locus $Q$ of A, D, or E orbifold
singularities. Many such examples arise by duality with the
heterotic string, but there may also be examples that have this
structure but are not dual in this way to heterotic string
compactifications. This mechanism of generating gauge symmetry
from $G_2$ manifolds has been discussed in \cite{acharya} and elsewhere.

\def\vol{{\rm vol}}
In analyzing the four-dimensional effective gauge and
gravitational couplings, the volumes of $Q$ and of $X$, which we
denote as $V_Q$ and $V_X$, will be important.  We will meet the
dimensionless number 
\begin{equation}\label{plon}
{a=\frac{V_X}{ V_Q^{7/3}}.}\end{equation}
 In models
with K3 fibrations, we have $V_X=V_{{\rm K3}}\cdot V_Q$. (Volumes
are multiplicative in fibrations such as $X\to Q$.) In the region
in which one can see the duality with the heterotic string (so
that the manifold $X$ definitely exists), $Q$ is very large
compared to the ${\bf T}^3$ or K3, and hence $a<<1$. Another
class of models arises from Type II orientifolds {\cite{cvetic}
-\cite{blum}}; in this case, the radius of the $M$-theory circle is a
factor in the volume of $X$ but not of $Q$, so again $a<<1$ when
the duality can be used to deduce the existence of $X$. It seems
reasonable to guess that at least for many or most $G_2$ manifolds
obtained by these dualities and useful  for phenomenology, there
is an upper bound of order one on the possible values of $a$,
with a singularity developing if one tries to make $a$ too
large.  This would be analogous to what happens in the strongly
coupled heterotic string if, for most values of the instanton
numbers at the two ends, one tries to make the length of the
eleventh dimension too long compared to the scale of the other
compact dimensions. Unfortunately, we have no way to prove our
conjecture that $a$ is generally bounded above or to determine
the precise bound.

%
%
%
%
%
One can also construct $G_2$ orbifolds in which it is possible to
have any value of $a$. (They lack the singularities, discussed
momentarily, that generate chiral fermions.) Also, in a local
picture , explicit $G_2$ metrics with $Q=\S^3$ or a finite
quotient of $\S^3$ are known {\cite{bryant},\cite{gibbons}}, and have
been studied in a number of recent papers
{\cite{amv},\cite{achar},\cite{pwitten},\cite{friedmann}}. (These can be used to get
gauge symmetry but not chiral fermions; symmetry breaking by
Wilson lines was incorporated in the last of those papers.) Those
papers focused on the behavior as $Q$ shrinks to a point,
whereupon topology-changing transitions may occur, as first
proposed in \cite{amv}.  When $Q$ shrinks to a point, a singularity
develops that is a cone on $\S^3\times \S^3$, or a finite
quotient.  It is not known that such a singularity can develop in
a compact $G_2$ manifold $X$, but presumably this is possible. In
an example in which this can occur, there is no upper bound on
$a$, but there may instead be a lower bound on $a$!

Since the heterotic string on a Calabi-Yau manifold (or a Type II
orientifold) can readily have chiral fermions,  it must also be
possible to get chiral fermions in $M$-theory on a $G_2$
manifold.   The precise mechanism was determined in \cite{bwitten}:
chiral multiplets are supported at points on $Q$ at which $X$ has
a singularity that is worse than an orbifold singularity.  At
these points, $Q$ itself is smooth but the normal directions to
$Q$ in $X$ have a singularity more complicated than the orbifold
singularity (\ref{loppy}). The relevant singularities have been studied
in {\cite{pwitten} - \cite{gb}}.  Their details will not be important in
the present paper.

\bigskip\noindent{\it Model Of Grand Unification}

In constructing a model similar to a four-dimensional grand
unified model,  we start with $M$-theory on ${\bf R}^4\times X$,
where ${\bf R}^4$ is four-dimensional Minkowski space and $X$ has
$G_2$ holonomy.  We assume that $X$ contains a three-manifold $Q$
with a $\Z_5$ orbifold singularity (described locally by (\ref{loppy}))
in the normal direction. This means  that $SU(5)$ gauge fields
propagate on ${\bf R}^4\times Q$.

We further assume that the first Betti number of $Q$ vanishes,
$b_1(Q)=0$. This is actually true in all known examples (and
notably it is true for examples arising by duality with the
heterotic string; in that context, $b_1(Q)=0$, since the first
Betti number of the Calabi-Yau threefold $Z$ is actually zero).
This means that, in expanding around the trivial $SU(5)$
connection on $Q$, there are no zero modes for gauge fields.  Such
zero modes would lead to massless chiral superfields in the
adjoint representation of $SU(5)$.

Instead, we assume that that there is a nontrivial finite
fundamental group $\pi_1(Q)$ and first homology group $H_1(Q)$. A
typical example (which can arise in duality with the heterotic
string) is a lens space, $Q=\S^3/\Z_q$ for some $q$.  We will keep
this in mind as a concrete example, for which we will give a
completely explicit formula for the threshold corrections. Having
a finite and nontrivial first homology makes it possible to break
$SU(5)$ to the standard model subgroup $SU(3)\times SU(2)\times
U(1)$ by a {\it discrete} choice of flat connection in the vacuum.
Just as in the case of the heterotic string on a Calabi-Yau
manifold \cite{chsw}, the discreteness leads to interesting
possibilities for physics, including options for solving the
doublet-triplet splitting problem.

For example, if $\pi_1(Q)=\Z_q$, we take the holonomy around a
generator of $\pi_1(Q)$ to be 
\begin{equation}\label{tumgo}
{U=\left(\begin{array}{ccccc}e^{4\pi i
w/q} & & & & \cr
     & e^{4\pi i
w/q}& & & \cr
     & &e^{4\pi i
w/q} & & \cr
     & & & e^{-6\pi i
w/q}& \cr
     & & & &e^{-6\pi i
w/q} \cr\end{array}\right)}\end{equation}
 with some integer $q$.  This breaks $SU(5)$ to
the standard model subgroup as long as $5w$ is not a multiple of
$q$.

For a more realistic GUT-like model, we also need various chiral
superfields.  These include Higgs bosons transforming  in the
${\bf 5}$ and $\overline {\bf 5}$ of $SU(5)$, as well as quarks
and leptons transforming as three copies of $\overline{\bf
5}\oplus {\bf 10}$, and possibly additional fields in real
representations of $SU(5)$. Such chiral superfields are localized
at points $P_i$ on $Q$ at which the singularity in the normal
direction is more severe than the orbifold singularity (\ref{loppy}).

Doublet-triplet splitting can be incorporated by assuming suitable
discrete symmetries.  In the example considered in \cite{gwitten}, $Q$
was the lens space $\S^3/\Z_q$, and the discrete symmetry group
was a copy of $F=\Z_n$ acting on $Q$.  The fixed point set of $F$
consisted of two circles, and by suitable assignments of the
points $P_i$ to the two circles, doublet-triplet splitting was
ensured.  We will not recall the details here as we will not make
explicit use of them. 
%
(For another approach to making the proton sufficiently long-lived, see
{\cite{FaraggiDW}}.)

\noindent{\it Kaluza-Klein Reduction}

Now we consider the Kaluza-Klein reduction of the vector
supermultiplet on ${\bf R}^4\times Q$ to give massive particles on
${\bf R}^4$.

We begin with the gauge field $A$.  It can be expanded around the
Wilson loop background $A_{cl}$ as  
\begin{equation}\label{olon}
{A=A_{cl}+ a_\mu
dx^\mu +\phi_\alpha dy^\alpha,}\end{equation}
 where $x^\mu$, $\mu=1,\dots,4$ are
coordinates on ${\bf R}^4$, and $y^\alpha$, $\alpha=1,\dots,3$ are
coordinates on $Q$. Here $a_\mu$ transforms as a gauge field on
${\bf R}^4$ and a scalar field on $Y$, and $\phi_\alpha$
transforms as a scalar on ${\bf R}^4$ and a one-form on $Y$. Both
$a_\mu$ and $\phi_\alpha$ are functions of both the $x$'s and the
$y$'s, and thus can be expanded 
\begin{equation}\label{tonon}
{\begin{aligned}a_\mu(x,y) &
=\sum_n a_\mu^{(n)}(x)\chi^{(n)}(y) \cr \phi_\alpha(x,y) & =\sum_m
\phi^{(m)}(x)\psi_\alpha^{(m)}(y). \cr\end{aligned}}\end{equation}
 Here $\chi^{(n)}$ and
$\psi^{(m)}$ are  eigenfunctions, respectively, of $\Delta_0$ and
$\Delta_1$, where by $\Delta_k$ we mean  the Laplacian acting on
$k$-forms with values in the adjoint representation of $SU(5)$:

\begin{equation}\label{jufo}
\begin{aligned} \Delta_0 \chi^{(n)} & = \lambda_n^{(0)}
\chi^{(n)}\cr \Delta_1\psi^{(m)} & =
\lambda_m^{(1)}\psi^{(m)}.\cr\end{aligned}\end{equation}
 The $k$-form eigenvalues
$\lambda^{(k)}_n$, $k=0,1$, are interpreted in four dimensions as
the mass squared.  The $a_\mu^{(n)}$ are vector fields in four
dimensions, and form part of a vector multiplet, while the
$\phi^{(n)}$ are scalars and form part of a chiral multiplet.

Along with the gauge field, the seven-dimensional vector multiplet
contains three additional scalar fields, which, in
compactification on ${\bf R}^4\times Q$, behave as another
one-form $\tilde\phi_\alpha$ on $Q$. As explained in \cite{acharya},
this happens because of the twisting of the normal bundle. $\tilde
\phi_\alpha$ has a similar Kaluza-Klein expansion: 
\begin{equation}\label{juggo}
{
\tilde\phi_\alpha(x,y) =\sum_m \tilde
\phi^{(m)}(x)\psi_\alpha^{(m)}(y).}\end{equation}
 The fields $\phi^{(m)}$ and
$\tilde \phi^{(m)}$ together make up the bosonic part of a chiral
multiplet.

In sum, then, each zero-form eigenfunction $\chi^{(n)}$ leads to a
vector multiplet, with helicities $1,-1,1/2,-1/2$, and each
one-form eigenfunction $\psi^{(m)}$ leads to a chiral multiplet,
with helicities $0,0,1/2,-1/2$.  (The origin of the fermionic
states in these multiplets is described in \cite{acharya}\ and depends
again on the twisting of the normal bundle.)  This is all the
information we need to compute the threshold corrections in
section 3.  However, we pause to describe in more detail the
Higgs mechanism, by virtue of which some chiral and vector
multiplets combine into massive vector multiplets.

If $\chi^{(n)}$ is a zero mode of the Laplacian, then
$a_\mu^{(n)}$ is a generator of the unbroken $SU(3)\times
SU(2)\times U(1)$. If $a_\mu^{(n)}$ is one of these 12 modes,
then it is part of a massless vector multiplet. For all the
infinitely many other values of $n$, $a_\mu^{(n)}$ is part of a
massive vector multiplet, obtained by combining a vector
multiplet and a chiral multiplet, via a Higgs mechanism.

Let us see how this comes about.    Since we have assumed that
$b_1(Q)=0$, there are no harmonic one-forms. Thus, any one-form
$\psi$ on $Q$ is a linear combination of a closed one-form
$d_A\chi$ and a co-closed one-form $d_A^*\lambda$ ($d_A$ and
$d_A^*$ are the gauge-covariant exterior derivative and its
adjoint, defined using the background gauge field $A_{cl}$; $\chi$
is a zero-form and $\lambda$ is a two-form). In particular, the
eigenfunctions $\psi^{(m)}$ are either closed or co-closed. The
co-closed eigenfunctions yield massive chiral multiplets in four
dimensions that do not participate in any Higgs mechanism. But a
closed one-form eigenfunction is the exterior derivative of a
zero-form eigenfunction, 
\begin{equation}\label{udni}
{\psi^{(m)}=d_A\chi^{(n)},}\end{equation}
 for
some $n$.  Here $\chi^{(n)}$ is a zero-form eigenfunction of the
Laplacian with the same eigenvalue as $\psi^{(m)}$. In this
situation, the vector multiplet derived from $\chi^{(n)}$ and the
chiral multiplet derived from $\psi^{(m)}$ combine via a Higgs
mechanism into a massive vector multiplet. The Higgs mechanism is
most directly seen using the transformation of the gauge field $A$
under a gauge transformation generated by the zero-form
$\chi^{(n)}$. This is $\delta A=d_A\chi^{(n)}$. In terms of
four-dimensional fields, it becomes 
\begin{equation}\label{ucx}
{\delta
a_\mu^{(n)}=\frac{D}{ Dx^\mu}\chi^{(n)}, ~~ \delta
\phi^{(m)}=\chi^{(n)},}\end{equation}
 where in the last formula, the shift in
$\phi^{(m)}$ under a gauge transformation demonstrates the Higgs
mechanism.

For our purposes, a vector multiplet, massless or not, has four
states of helicities $\pm 1,\pm 1/2$, and a chiral multiplet has
four states of helicities $0,0,\pm 1/2$.  A massive vector
multiplet is the combination of a vector multiplet and a chiral
multiplet via the Higgs mechanism to a multiplet with eight
helicity states $1,1/2,1/2,0,0,-1/2,-1/2,-1$.  We will not use
the concept of a massive vector multiplet in the rest of the
paper, since for computing the threshold corrections, one need
not explicitly take account of the way the Higgs effect combines
a vector multiplet and a chiral multiplet into a single massive
vector multiplet.

For more detail, with also some information that will be useful
later, see the table.  As we note in the table, massless states
come from harmonic forms (in the above discussion and in the rest
of the paper, we make topological restrictions on $Q$ so that
there are no harmonic one-forms leading to massless chiral
multiplets, but in the table we include them for completeness and
possible future generalizations).  We have also indicated in the
table which massive multiplets are Higgsed and become part of
massive vector multiplets, and which are not.

\begin{table}
\caption{Field content in four dimensions.}
\vspace{.25em}
\begin{small}
\begin{tabular}{|@{\hspace{.3em}}c@{\hspace{.3em}}|@{\hspace{.3em}}c
    @{\hspace{.3em}} 
    |@{\hspace{.3em}}c@{\hspace{.3em}}|@{\hspace{.3em}}c@{\hspace{.3em}}
    |@{\hspace{.3em}}c@{\hspace{.3em}}|@{\hspace{.3em}}c@{\hspace{.3em}}|} 
\hline \hline
        &Form on $Q$   & Type
        &Supermultiplet & $\chi $ &Str $ (\frac{1}{ 12}-\chi ^2)$ \\\hline\hline
massive &0--forms&& vector Higgsed& $1,1/2,-1/2,-1$&$-3/2$ \\\hline
   ''     & 1--forms & closed & chiral Higgsed    & $1/2,0,0,-1/2$& $1/2$ \\\hline
     ''   &    ''    & co-closed & chiral un-Higgsed & $1/2,0,0,-1/2$&
        $1/2$\\ \hline \hline
massless&0--forms &harmonic & vector&$1,1/2,-1/2, -1$& $-3/2$\\\hline
     ''  &1--forms &harmonic& chiral & $1/2,0,0,-1/2$&$1/2$ \\ \hline
\end{tabular}
\end{small}
\end{table}


\section{The Threshold Corrections}
\label{sec:3}

In this section, we will get down to business and compute the
threshold corrections to gauge couplings in grand unification. In
doing the computation, it is necessary to assume that the radius
$R_Q$ of $Q$ is much greater than the eleven-dimensional Planck
length $1/M_{11}$, since otherwise $M$-theory is strongly coupled
at the scale $1/R_Q$.  Having $R_Q>>1/M_{11}$ also means that the
Kaluza-Klein harmonics, with masses of order $M_{GUT}\sim 1/R_Q$,
are much lighter than generic $M$-theory excitations, whose
masses are of order $M_{11}$.  So it makes sense to compute their
effects without knowing exactly what happens at the Planck scale.
In section 4, we show that $R_Q>>1/M_{11}$ to the extent that
$\alpha_{GUT}$ is small.

In most of this section, we consider only the case that the normal
bundle to $Q$ has only the standard orbifold singularity, so that
the only charged light fields are those of the seven-dimensional
vector multiplet, compactified on $Q$. After analyzing this case
thoroughly, we consider in section 3.5 the case in which
singularities of the normal bundle generate additional light
charged fields.

{\it A priori}, one would expect the one-loop threshold
corrections to gauge couplings to depend on the chiral multiplets
that describe the moduli of $X$.  However, the imaginary parts of
those multiplets are axion-like components of the $C$-field and
manifestly decouple from the computation of one-loop threshold
corrections.\footnote{Nonperturbatively small threshold corrections
due to membrane instantons will depend on the $C$-field.  These
effects, though extremely small for $G_2$ manifolds, can be
substantial in a dual heterotic string or Type II orientifold
description.} Those computations depend only on the particle
masses -- that is, the eigenvalues of the Laplacian. Hence the
perturbative threshold corrections are actually constants,
independent of the moduli of $X$.

The threshold corrections are given by a sum over contributions of
different eigenvalues. The sum makes sense for any metric on $Q$,
but  we want to evaluate it for metrics that are induced by $G_2$
metrics on $X$. From the argument in the last paragraph, the
threshold correction is independent of the metric on $Q$ as long
as it comes from a $G_2$ metric on $X$.  We do not know any useful
property of such metrics on $Q$. So the most obvious way for the
threshold correction to be independent of the moduli of $X$ is for
it to be entirely independent of the metric of $Q$, that is, to be
a topological invariant.

The most obvious topological invariant of a three-manifold $Q$,
endowed with a background flat gauge field $A$, that can be
computed from the spectrum of the Laplacian is the Ray-Singer
analytic torsion \cite{rs}.  We will show that torsion does indeed give
the right answer.

{}From a physical point of view, the torsion can be represented by
a very simple topological field theory \cite{aschwarz}. Another hint that torsion is
relevant is that analytic torsion (in this case, $\bar\partial$
analytic torsion, which varies holomorphically and is not a
topological invariant) governs the threshold corrections for the
heterotic string, as shown in section 8.2 of \cite{bcos}.  The relation of the threshold corrections for
$G_2$ manifolds to torsion can possibly also be argued using
methods in \cite{bcos}, at least in the case of gauge groups of type A
or D, plus duality of $M$-theory with Type IIA superstrings and
$D$-branes. We will establish this relation directly.  (Torsion
also enters in the theory of membrane instantons \cite{hmt}.)

\def\R{{\cal R}}
\def\K{{\cal K}}
\def\T{{\cal T}}
\subsection{The Analytic Torsion}
\label{sec:3.1}

Let $H$ be the subgroup of the unified group $G$ that commutes
with the standard model group $SU(3)\times SU(2)\times U(1)$. For
$G=SU(5)$, $H$ is just the $U(1)$ of the standard model. Let
$\R_i$ be the standard model representations appearing in the
adjoint representation of $G$, and suppose that the part of the
adjoint representation of $G$ transforming as $\R_i$ under the
standard model transforms as $\omega_i$ under $H$.  (Some of the
$\omega_i$ may be the same for different $\R_i$; also, the
$\omega_i$ are not necessarily irreducible, though they are
irreducible for $G=SU(5)$.)

In vacuum, we suppose that $G$ is broken to the standard model by
a choice of a flat $H$-bundle on $Q$, that is, by a choice of
Wilson lines. Each $\omega_i$ determines a flat bundle that we
will denote by the same name.  We want to define the analytic
torsion $\T_i$ for the representation $\omega_i$. Let
$\Delta_{k,i}$, $k=0,\dots,3$, be the Laplacian acting on
$k$-forms with values in $\omega_i$. If there is no cohomology
with values in $\omega_i$, that is, if the $\Delta_{k,i}$ have no
zero modes, then the torsion is defined by the formula\footnote{We
define the torsion as in \cite{rs}\ and much of the physics literature;
the definition differs by a factor of $-2$ from that in \cite{ray}.  We
also note that what we call $e^\T$ is called the torsion by many
authors. }
\begin{equation}\label{inuncu}
{\T_i=\frac{1}{ 2}\sum_{k=0}^3
(-1)^{k+1}k\log\det\left(\Delta_{k,i}/\Lambda^2\right),}\end{equation}
where
$\Lambda$ is an arbitrary constant (which in our physical
application will be the gauge theory cutoff). The determinants in
(\ref{inuncu})\ are defined using zeta function regularization (for more
detail on this, see Appendix A where a simple example is
computed). By the theorem of Ray and Singer, $\T_i$ is independent
of the metric of $Q$, and hence also independent of $\Lambda$,
which can be eliminated by scaling the metric of $Q$. (In the
mathematical theory, $\Lambda$ is usually just set to 1, but
physically, since $\Delta_{k,i}$ is naturally understood to have
dimensions of mass squared, we introduce a mass parameter
$\Lambda$ and use the dimensionless ratio
$\Delta_{k,i}/\Lambda^2$.) Because the Laplacian commutes with the
Hodge $*$ operator which maps $k$-forms to $(3-k)$-forms,
$\Delta_{k,i}$ and $\Delta_{3-k,i}$ have the same spectrum. So we
can simplify (\ref{inuncu}): 
\begin{equation}\label{torsionu}
{\T_i=\frac{3}{
2}\log\det\left(\Delta_{0,i}/\Lambda^2\right)-\frac{1}{
2}\log\det\left(\Delta_{1,i}/\Lambda^2\right).}\end{equation}

If $\omega_i$ is such that the $\Delta_{k,i}$ do have zero modes,
then the $\log\det$'s in (\ref{inuncu})\ are not well-defined.  In this
more general case, we define objects $\K_i$ by replacing
$\det\Delta_{k,i}$ (which vanishes when there are zero modes) with
$\det{}'\Delta_{k,i}$, the product (regularized via zeta
functions) of the non-zero eigenvalues of $\Delta_{k,i}$:
\begin{equation}\label{oorsionu}
{\K_i=\frac{3}{
2}\log\det{}'\left(\Delta_{0,i}/\Lambda^2\right)-\frac{1}{
2}\log\det {}'\left(\Delta_{1,i}/\Lambda^2\right).}\end{equation}
 Since
$\log\det$ is the same as $\Tr\log$, we can equally well write
\begin{equation}\label{orsionu}
{\K_i=\frac{3}{
2}\Tr'\log\left(\Delta_{0,i}/\Lambda^2\right)-\frac{1}{
2}\Tr'\log\left(\Delta_{1,i}/\Lambda^2\right),}\end{equation}
 where $\Tr'$ is a
trace with zero modes omitted.

\def\00{{\cal O}}
\def\K{{\cal K}}
For a three-manifold $Q$ with finite fundamental group, and a
non-trivial irreducible representation $\omega_i$, there are no
zero modes. If $\omega_i$ is trivial, there are zero modes: they
are simply the zero-form 1 and a three-form which is the
covariantly constant Levi-Civita volume form. When there are zero
modes, $\T_i$ does not simply equals $\K_i$; there is a correction
for the zero modes. The correction is explained in Appendix B, and
in the present situation is very simple.  If $V_Q$ is the volume
of $Q$, then the torsion of the trivial representation (which we
denote by ${\cal O}$) is
\begin{equation}\label{pinuncu}
{\T_{\cal O}
={\cal K}_{\cal O}-\log(V_Q\Lambda^3).}\end{equation}
  The Ray-Singer theorem
again implies that $\T_\00$ is independent of the metric of $Q$,
and hence in particular independent of $\Lambda$.  In the
mathematical literature, (\ref{pinuncu})\ would be written with
$\Lambda=1$. In general, for any irreducible representation
$\omega_i$, trivial or not,
\begin{equation}\label{inuncu}
{\T_i=\K_i-\delta_{\omega_i,{\cal O}}
\log(V_Q\Lambda^3).}\end{equation}

\subsection{Sum Over Massive Particles}
\label{sec:3.2}

Now we will describe the threshold correction. We let $g_M$ be the
underlying gauge coupling as deduced from $M$-theory in the
supergravity approximation (we compute it in detail in section 4),
and we let $g_a(\mu)$, $a=1,2,3$ be the standard model $U(1)$, $
SU(2)$, and $ SU(3)$ gauge couplings measured at an energy $\mu$,
which we assume to be far below the cutoff $\Lambda$.

The tree level relations among the $g_a$ depend on how they are
embedded in $G$.  For the usual embedding\footnote{Conventionally,
$SU(2)$ and $SU(3)$ are generated by $5\times 5$ matrices of
trace-squared equal to $1/2$, and $U(1)$ is generated by the
hypercharge, ${\rm diag}(1/2,1/2,-1/3,-1/3,-1/3)$, whose
trace-squared is $(5/3)\cdot 1/2$.} of the standard model in
$SU(5)$, the tree level relations are $g_a^2=g_M^2/k_a$, with
$(k_1,k_2,k_3)=(5/3,1,1)$.

\def\S{{\cal S}}
\def\Str{{\,{\rm Str}\,}}
The one-loop relation is\footnote{The threshold corrections are called
$\Delta_a$ in \cite{kaplunovsky}, but we will call them $\S_a$ to avoid
confusion with the use of $\Delta$ for the Laplacian.}
\begin{equation}\label{turmigo}
{\frac{16\pi^2}{ g_a^2(\mu)}
=\frac{16\pi^2k_a}{ g_M^2}+b_a\log(\Lambda^2/\mu^2)+\S_a,}\end{equation}
 where
$b_a$ are the one-loop beta function coefficients, and $\S_a$ are
the one-loop threshold corrections. They are given by very similar
sums over, respectively, the massless and massive states. The
formula for $b_a$ is 
\begin{equation}\label{hoco}
{b_a=2\Str_{m=0} Q_a^2\left(\frac{1}{
12}-\chi^2\right).}\end{equation}
 Here $\Str$ is a supertrace (bosons contribute
with weight $+1$, fermions with weight $-1$) over massless
helicity states; $\chi$ is the helicity operator; and $Q_a$ is a
generator of the $a^{th}$ factor of the standard model group with
$\Tr\,Q_a^2=k_a/2$.  As explained in the introduction to this
section, until section 3.5 we consider the effects of the
seven-dimensional vector multiplets only.  In this case, the
massless particles are simply the vector multiplets of the
unbroken group $SU(3)\times SU(2)\times U(1)$.  So, letting ${\cal
R_A}$ denote the adjoint representation of the standard model, and
recalling that the helicities of a vector multiplet are
$1,-1,1/2,-1/2$, we see that for this case,
\begin{equation}\label{loco}
{b_a=-3\Tr_{\cal R_A}Q_a^2.}\end{equation}

The definition of $\S_a$ is rather similar to (\ref{hoco})\ except that
the trace runs over massive states and includes a logarithmic
factor depending on the mass:
\begin{equation}\label{hhoco}
{\S_a=2\Str_{m\not=0}
Q_a^2\left(\frac{1}{ 12}-\chi^2\right)\log(\Lambda^2/m^2).}\end{equation}

If we combine these formulas, we can write
\begin{equation}\label{urmigo}
\begin{aligned}\frac{16\pi^2}{ g_a^2(\mu)} =\frac{16\pi^2k_a}{
g_M^2}&+2\Str_{m=0} Q_a^2\left(\frac{1}{
12}-\chi^2\right)\log(\Lambda^2/\mu^2)\cr & +2\Str_{m\not=0}
Q_a^2\left(\frac{1}{ 12}-\chi^2\right)\log(\Lambda^2/m^2).\cr\end{aligned}\end{equation}
 We
see that every helicity state, massless or massive, makes a
contribution to the low energy couplings that has the same
dependence on the cutoff $\Lambda$, independent of its mass.

In a unified four-dimensional GUT theory, the quantum numbers of
the tree level particles are $G$-invariant, though the masses are
not.  Hence, the coefficient of $\log\Lambda$ in (\ref{urmigo})\ is
$G$-invariant.  The precise value of $\Lambda$ is therefore
irrelevant in the sense that a change in $\Lambda$ can be
absorbed in redefining the unified coupling (which in a
four-dimensional theory would usually be called $g_{GUT}$ rather
than $g_M$).

In our supersymmetric gauge theory on ${\bf R}^4\times Q$, the sum
(\ref{urmigo})\ arises from a one-loop diagram in the seven-dimensional
gauge theory; the coupling renormalization from the one-loop
diagram has been expanded as a sum over Kaluza-Klein harmonics.
The seven-dimensional gauge theory is unrenormalizable, and
divergences in loop diagrams should be expected.  The divergences,
however, are proportional to gauge-invariant local operators on
$Q$.  Since we have assumed that the $G$-symmetry is broken only
by the choice of a background flat connection, all local operators
are $G$-invariant.  Hence the divergences are $G$-invariant and
can be, again, absorbed in a redefinition of $g_M$.

In practice (by using the relation to analytic torsion), we will
define the sum in (\ref{urmigo})\ with zeta-function regularization, and
then it will turn out that there are no divergences at all --
$\Lambda$ will cancel out completely.  With a different
regularization, there would be divergences, but they would be
$G$-invariant.  Of course, if we really had a proper understanding
of $M$-theory, we would  use whatever regularization it gives.

\subsection{Relation Of The Threshold Correction To The Torsion}
\label{sec:3.3}

The threshold correction $\S_a$ can be written
$\S_a=\sum_i\S_{a,i}$, where $\S_{a,i}$ is the contribution to the
sum in (\ref{hhoco})\ coming from states that transform in the
representation $\R_i$.  We can further factor the trace of $Q_a^2$
-- which depends only on $\R_i$ -- from the rest of the sum:
\begin{equation}\label{invo}
{\S_{a,i}=2\Tr_{\R_i}Q_a^2\,\Str_{{\cal
H}_i}\left(\frac{1}{ 12}-\chi^2\right)\log\left(\frac{\Lambda^2 }
{m^2}\right).}\end{equation} Here ${\cal H}_i$ is defined by saying that the space
${\cal H}$ of one-particle massive states decomposes under the
standard model as ${\cal H}=\oplus \R_i\otimes {\cal H}_i$.

Now let us evaluate the sum over helicity states.  As we reviewed
in section 2,  each eigenvector of the zero-form Laplacian
$\Delta_{0,i}$ contributes to ${\cal H}_i$ a vector multiplet,
with helicities $1,-1,1/2,-1/2$.  $m^2$ for this multiplet is
equal to the eigenvalue of $\Delta_{0,i}$.  For such a multiplet,
we have $\Str\left(\frac{1}{ 12}-\chi^2\right) = -3/2$. Each
eigenfunction of the one-form Laplacian $\Delta_{1,i}$ similarly
contributes to ${\cal H}_i$ a chiral multiplet, with helicities
$0,0,1/2,-1/2$, for which $\Str\left(\frac{1}{
12}-\chi^2\right)=1/2$. Inserting these values in (\ref{invo})\ and using
(\ref{orsionu})\ and (\ref{inuncu}), we find that
\begin{equation}\label{pinvo}
{\S_{a,i}=2\Tr_{\R_i}Q_a^2\cdot\K_i=2\Tr_{\R_i}Q_a^2\cdot
\left(\T_i +\delta_{\omega_i,{\cal O}}\log(V_Q\Lambda^3)\right).}\end{equation}

Here $\delta_{{\omega_i},{\cal O}}$ is 1 if $\omega_i$ is the
trivial representation (which we denote as ${\cal O}$) and
otherwise zero.\footnote{If it should happen that $\omega_i$ is
reducible for some $i$, then $\delta_{\omega_i,{\cal O}}$ should
be understood as the number of copies of ${\cal O}$ in $\omega_i$.
In what follows, we assume for ease of exposition that the
$\omega_i$ are all irreducible, as occurs for $G=SU(5)$.  This
restriction is inessential.}

\bigskip\noindent{\it Elimination Of $\Lambda$ Dependence}

Now we can eliminate the $\Lambda$ dependence, which appears
explicitly in (\ref{turmigo})\ in the contribution from the massless
particles, and arises in (\ref{pinvo})\ because massless contributions
are omitted in $\S_a$. The $\Lambda$ dependence of $1/g_a^2(\mu)$
cancels only when we include all states, massive and massless, and
we will now exhibit this cancellation.

The total $\Lambda$-dependent contribution to $\S_a$ is, using
(\ref{pinvo}), 
\begin{equation}\label{jinvo}
{2\sum_i \Tr_{\R_i}Q_a^2\delta_{\omega_i,{\cal
O}}\log(V_Q\Lambda^3).}\end{equation}
 We can write the adjoint representation
$\A$ of the low energy gauge group $SU(3)\times SU(2)\times U(1)$
as 
\begin{equation}\label{gufot}
{\A=\oplus'_i\R_i,}\end{equation}
 where the sum $\oplus_i'$ runs
over all $\R_i$ such that $\omega_i={\cal O}$. The reason for this
is
 that when we turn on a flat gauge background of a subgroup
$H\subset G$ to break the gauge group $G$ to a subgroup, the
unbroken group is the subgroup that transforms trivially under
$H$, in other words, its Lie algebra is the union of the $\R_i$
for which $\omega_i={\cal O}$.  With (\ref{gufot}), we can rewrite
(\ref{jinvo})\ as 
\begin{equation}\label{newjinvo}
{2\Tr_{\cal A}Q_a^2 \log(V_Q\Lambda^3),}\end{equation}
and this is the $\Lambda$ dependence of $\S_a$.

The $\Lambda$-dependence that is explicit in (\ref{turmigo})\ is in the
term $b_a\log(\Lambda^2/\mu^2)$. Using (\ref{loco}), we see that the
$\Lambda$ dependence cancels with that in
(\ref{newjinvo}), and moreover, making use of (\ref{pinvo})\ and
(\ref{newjinvo})\ as well as (\ref{loco}), we can rewrite the formula for the
threshold corrections in a $\Lambda$-independent and useful form:
\begin{equation}\label{torry}
{\frac{16\pi^2}{ g_a^2(\mu)} =\frac{16\pi^2k_a}{
g_M^2}+b_a\log\left(\frac{1}{ V_Q^{2/3}\mu^2}\right)+\S_a',}\end{equation}
with
\begin{equation}\label{jigo}
{\S_a'=2\sum_i \T_i\,\,\Tr_{\R_i} Q_a^2.}\end{equation}

\subsection{Evaluation For $SU(5)$}
\label{sec:3.4}

\def\8{{\bf 8}}
\def\1{{\bf 1}}
\def\3{{\bf 3}}
\def\2{{\bf 2}}
Now, let us evaluate this formula for $G=SU(5)$.  The adjoint
representation decomposes under the standard model  as
\begin{equation}\label{bronco}
{(\8,\1)^{0}\oplus (\1,\3)^{0}\oplus(\1,\1)^{0}\oplus
(\3,\2)^{-5/6}\oplus (\overline \3,\2)^{5/6},}\end{equation}
 where $SU(3)\times
SU(2)$ representations have been labeled mostly by their
dimension, and the superscript is the $U(1)$ charge. Thus, the
representations of $H=U(1)$ that enter are the trivial
representation ${\cal O}$, a nontrivial representation $\omega$
that corresponds to charge $-5/6$, and the dual representation
$\overline\omega$ for charge $5/6$.  Since  complex conjugation of
the eigenfunctions exchanges $\omega$ and $\overline \omega$
without changing the eigenvalues of the Laplacian, we have
$\T_\omega=\T_{\overline\omega}$.  Hence, the can write the
formula for $\S_a'$ just in terms of the two torsions $\T_\00$ and
$\T_\omega$.

To make this explicit, we need to take a few traces.  The traces
of $(Q_1^2,Q_2^2,Q_3^2)$ in the representation $(\8,\1)^{0}\oplus
(\1,\3)^{0}\oplus(\1,\1)^{0}$ are $(0,2,3)$, and their traces in
the representation $(\3,\2)^{-5/6}\oplus (\3,\overline \2)^{5/6}$
are $(25/3,3,2)$.  So 
\begin{equation}\label{forme}
\begin{aligned}\S_1' & =\frac{50}{
3}\T_\omega\cr \S_2' & = 4\T_\00+6\T_\omega \cr \S_3' & =
6\T_\00+4\T_\omega.\cr\end{aligned}\end{equation}
 Since $(b_1,b_2,b_3)=-3(0,2,3)$, we can
write this as 
\begin{equation}\label{onfor}
{\S_a'=-\frac{2}{
3}b_a\left(\T_\00-\T_\omega\right) +10 k_a\T_\omega.}\end{equation}
 So we get
our final formula for the low energy gauge couplings in the
one-loop approximation: 
\begin{equation}\label{finalform}
{\frac{16\pi^2}{
g_a^2(\mu)}=\left(\frac{16\pi^2}{ g_M^2}+10\T_\omega\right)k_a+b_a
\log\left(\frac{\exp\left(\frac{2}{ 3}(\T_\omega-T_\00)\right) }
{
\mu^2V_Q^{2/3}}\right).}\end{equation}

We might compare this to a naive one-loop renormalization group
formula that we might write in a GUT theory.  This would read
\begin{equation}\label{inalform}
{\frac{16\pi^2}{ g_a^2(\mu)}=\left(\frac{16\pi^2}{
g_{GUT}^2}\right)k_a+b_a \log(M_{GUT}^2/\mu^2).}\end{equation}
 We see that the
two formulas agree if 
\begin{equation}\label{truro}
\begin{aligned}\frac{16\pi^2}{
g_{GUT}^2}&=\frac{16\pi^2}{ g_{M}^2}+10\T_\omega\cr
             M_{GUT}^2&= \left(\frac{\exp(\T_\omega-\T_\00) }{
             V_Q}\right)^{2/3}.\cr\end{aligned}\end{equation}
The first formula tells us how the coupling $g_M$   used in
$M$-theory should be compared to the $g_{GUT}$ that is inferred
from low energy data.  Our computation really only makes sense if
the difference between $g_M$ and $g_{GUT}$ is much smaller than
either, since otherwise higher order corrections would be
important. Moreover, a different regularization (such as
$M$-theory may supply) might  have given a different answer for
the shift in $g_M$, so this shift is unreliable.
 The second formula shows how the parameter
$V_Q$ of the compactification is related to $M_{GUT}$ as inferred
from low energy data.\footnote{We define $M_{GUT}$ as the mass
parameter that appears in making a fit like (\ref{inalform})\ to low
energy data; it is of course not necessarily the mass of any
particle.}  This relationship is meaningful and independent of the
regularization.  We can write the relation as
\begin{equation}\label{timely}
{V_Q=\frac{L(Q)}{ M_{GUT}^3},~~{\rm
with}~L(Q)=\exp(\T_\omega-\T_\00).}\end{equation}

A noteworthy fact -- though a simple consequence of $SU(5)$ group
theory -- is that the massive Kaluza-Klein harmonics have made no
correction at all to the prediction of the theory for
$\sin^2\theta_W$.

\def\S{{\bf S}}
\def\Z{{\bf Z}}
We now want to make our result completely explicit in a simple
example. To allow for $SU(5)$ breaking, $Q$ cannot be the most
obvious compact three-manifold with $b_1=0$, which would be a
sphere $\S^3$.  We can, however, take $Q$ to be what is arguably
the next simplest choice, a lens space.  We describe $\S^3$ by
complex variables $z_1,z_2$ with $|z_1|^2+|z_2|^2=1$, and take
$\Z_q$ to act by 
\begin{equation}\label{unu}
{\gamma:z_i\to \exp(2\pi i/q)z_i}\end{equation}
 for
some positive integer  $q$. Then we define 
\begin{equation}\label{tuno}
{Q=\S^3/\Z_q.}\end{equation}
To break $SU(5)$ to the standard model, we assume that the action
of $\gamma$ is accompanied by a gauge transformation by
\begin{equation}U_\gamma=\exp\left(2\pi i (w/q){\rm
    diag}(2,2,2,-3,-3)\right)\end{equation} 
with some integer $w$ such that $5w$ is not divisible by $q$. The
torsions in this case are 
\begin{equation}\label{gudu}
{\T_\00=-\log q,~~\T_\omega=
\log\left(4\sin^2(5\pi w/q)\right),}\end{equation}
 giving
\begin{equation}\label{spico}
{L(Q)=4q\sin^2(5\pi w/q).}\end{equation}
(See {\cite{ray},\cite{rs}}
as well as Appendix A for some computations.)   Hence the relation
between $M_{GUT}$ and $V_Q$ is in this model
\begin{equation}\label{relq}
{M_{GUT}=\left(\frac{4q\sin^2(5\pi w/q)}{
V_Q}\right)^{1/3}, ~~ V_Q=\frac{4q\sin^2(5\pi w/q)}{ M_{GUT}^3}.}\end{equation}

We actually can generalize (\ref{unu})\ slightly to a transformation
$\gamma$ that maps $z_i\to \exp(2\pi im_i/q)z_i$ for integers
$m_i$ that are prime to $q$.  The quotient $\S^3/\Z_q$ is still
called a lens space.  By replacing $\gamma$ by a power of itself,
there is no loss of generality to take, say $m_2=1$; we then
denote $m_1$ simply as $m$. In this more general case (see \cite{rs},
pp. 168-9), $\T_\00$ is unchanged, and $\T_\omega$ becomes
$\log(4|\sin(5\pi w/q)\sin(5\pi jw/q)|)$ where $jm\equiv 1$ modulo
$q$, so $L(Q)=4q|\sin(5\pi w/q)\sin(5\pi jw/q)|$.

\subsection{Inclusion Of Quarks, Leptons, And Higgs Bosons}
\label{sec:3.5}

So far we have solely considered  the case that the normal space
to $Q$ has only the standard orbifold singularity, so that the
only charged particles with masses $\leq 1/R_Q$ are the
Kaluza-Klein harmonics.  Now we want to introduce quarks, leptons,
Higgs bosons, and possibly other charged light fields such as
messengers of gauge-mediated supersymmetry breaking.  We do this
by assuming at points $P_i\in Q$ the existence of certain more
complicated singularities of the normal bundle.  These generate
charged massless $SU(5)$ multiplets (which may ultimately get
masses at a lower scale if a superpotential is generated or
supersymmetry is spontaneously broken). If the singularities of
the $P_i$ are generic, each one contributes a new irreducible
$SU(5)$ multiplet $M_i$ of massless chiral superfields.
 Specific singularities that generate
  chiral multiplets transforming in the ${\bf 5}$, ${\bf 10}$,
$\overline {\bf 10}$, and $\overline {\bf 5}$ of $SU(5)$ have been
studied in {\cite{pwitten} - \cite{gb}}.

 It is believed that these singularities are
{\it conical}. This is definitely true in a few cases in which the
relevant $G_2$ metrics are conical metrics that were constructed
long ago {\cite{bryant},\cite{gibbons}} (and found recently \cite{pwitten}\ to
generate massless chiral multiplets). Since a conical metric
introduces no new length scale that is positive but smaller than
the eleven-dimensional Planck length, we expect that these
singularities, apart from the massless multiplets $M_i$, introduce
no new particles of masses $\leq 1/R_Q$ that need to be considered
in evaluating the threshold corrections.

We can also argue, a little less rigorously, that the
singularities of the normal bundle that produces massless chiral
superfields in the ${\bf 5}$, ${\bf 10}$, etc., have no effect on
the Kaluza-Klein harmonics of the seven-dimensional vector
multiplet on ${\bf R}^4\times Q$.  To show this, we consider the
construction in \cite{bwitten}, where the association of massless chiral
multiplets with singularities was argued using duality with the
heterotic string. In this argument, the existence in the $G_2$
description of a conical singularity  that generates massless
chiral superfields was related, in a heterotic string description
that uses a ${\bf T}^3$ fibration, to the existence of a certain
zero mode of the Dirac equation on a special ${\bf T}^3$ fiber.
The existence of this zero mode generates a localized massless
multiplet in the ${\bf 5}$, ${\bf 10}$, etc., as shown in
\cite{bwitten}, but does nothing at all to the seven-dimensional vector
multiplet (which has no exceptional zero mode on the ${\bf T}^3$
in question).

Granted these facts, to incorporate the effects of the multiplets
$M_i$, all we have to do is add their contributions to the
starting point (\ref{turmigo})\ or to the final result (\ref{finalform}). If all
of the new multiplets are massless down to the scale of
supersymmetry breaking, then, for $\mu$ greater than this scale,
all we have to do is add a contribution to (\ref{turmigo})\ due to the
new light fields.  Let $\Delta b_a$ be the contribution of the new
light fields to the beta function coefficients $b_a$ -- note that
since the $M_i$ form complete $SU(5)$ multiplets, they contribute
 to each $b_a$ in proportion to $k_a$.  The contribution of the
new fields to (\ref{turmigo})\ is to add to the right hand side
\begin{equation}\label{hormigo}
{\Delta b_a\log(\widetilde \Lambda^2/\mu^2),}\end{equation}
 which is
the contribution due to renormalization group running of the $M_i$
from their cutoff $\widetilde \Lambda$ down to $\mu$.  We do not
exactly know what effective cutoff $\widetilde\Lambda$ to use for
the $M_i$, but it is of order $M_{11}$.  Anyway, the exact value
of $\widetilde \Lambda$ does not matter; it can be absorbed in a
small correction to $g_M$ (this correction is no bigger than other
unknown corrections due for example to possible charged particles
with masses of order $M_{11}$).   In fact, up to a small shift in
$g_M$, it would not matter if we replace $\widetilde\Lambda$ by
the (presumably lower) mass $\exp\left(\frac{1}{ 3}
(\T_\omega-\T_\00)\right)/V_Q^{1/3}$ that appears in (\ref{finalform}).
So if all components of the $M_i$ are light, we can take our final
answer to be simply that of (\ref{finalform}), but with all $b_a$
redefined (by the shift $b_a\to b_a+\Delta b_a$) to include the
effects of the $M_i$.  In other words, if all components of $M_i$
are light, we simply have to take the $b_a$ in (\ref{finalform})\ to be
the exact $\beta$ function coefficients of the low energy theory.

The assumption that all components of the $M_i$ are light is
inconsistent with the measured value of the weak mixing angle
$\sin^2\theta_W$. That measured value (and the longevity of the
proton) is instead compatible with the hypothesis that all
components of the $M_i$ are light except for the color triplet
partners of the ordinary $SU(2)\times U(1)$ Higgs bosons; we call
those triplets $T$ and $\tilde T$. Let $m_T$ be the mass of $T$
and $\tilde T$ (we assume this mass comes from a superpotential
term $T\tilde T$, in which case $T$ and $\tilde T$ have equal
masses), and let $\Delta b_a^{T,\widetilde T}$ be their
contribution to the beta functions. These are not proportional to
$k_a$ since $T$ and $\tilde T$ do not form a complete $SU(5)$
multiplet!  Then (\ref{hormigo})\ should be replaced by 
\begin{equation}\label{pormigo}
{
(\Delta b_a-b_a^{T,\widetilde
T})\log(\widetilde\Lambda^2/\mu^2)+b_a^{T,\widetilde
T}\log(\widetilde\Lambda^2/m_T^2),}\end{equation}
 the idea being that the $T$,
$\widetilde T$ contributions run only from $\widetilde\Lambda$
down to $m_T$, while the others run down to $\mu$. Up to a small
correction to $g_M$, we can again replace $\widetilde\Lambda$  in
(\ref{pormigo})\ by 
\begin{equation}\label{ugu}
{\widetilde\Lambda\to \exp\left(\frac{1}{ 3}
(\T_\omega-\T_\00)\right)V_Q^{-1/3}.}\end{equation}
 If we do this, then
(\ref{finalform})\ is replaced by 
\begin{equation}\label{realfinalform}
\begin{aligned}
\frac{16\pi^2}{ g_a^2(\mu)}=&\left(\frac{16\pi^2}{ g_M^2}+10\T_\omega
+\delta\right)k_a +b_a \log\left(\frac{\exp(\frac{2}{
3}(\T_\omega-T_\00)) }{ \mu^2V_Q^{2/3}}\right)\cr &
+b_a^{T,\widetilde T}\log\left(\frac{\exp(\frac{2}{
3}(\T_\omega-T_\00)) }{ m_T^2V_Q^{2/3}}\right).\cr\end{aligned}\end{equation}
 Here $b_a$
are the full beta functions of the low energy theory below the
mass $m_T$, and $b_a^{T,\tilde T}$ is the additional contribution
to the beta functions from $T,\widetilde T$ between $m_T$ and the
effective GUT mass $M_{GUT}=\exp\left(\frac{1}{
3}(\T_\omega-\T_\00)\right)/V_Q^{1/3}$. Finally, $\delta$
expresses an unknown shift in the effective value of $g_M$; this
shift is presumably unimportant within the accuracy of the
computation.

Assuming that low energy threshold corrections are small,  the fit
to low energy measurements of gauge couplings is improved if
rather than $m_T\sim M_{GUT}$ we take 
\begin{equation}\label{guny}
{\frac{m_T}{
M_{GUT}}\sim 10^{-2}.}\end{equation}
 It is at least somewhat plausible in the
present model that $m_T/M_{GUT}$ would be small since the
superpotential term $m_TT\tilde T$ probably has to arise (like the
terms that lead to quark and lepton masses) from membrane
instantons.  We discuss this issue further in section 6.

\section{Couplings And Scales}
\label{sec:4}

One virtue of computing the threshold corrections is that we can
make somewhat more precise the formulas for the parameters
$M_{GUT}$, $\alpha_{GUT}$, and $G_N$ that are read off from the
eleven-dimensional supergravity action.

We write the gravitational action in eleven dimensions as
\begin{equation}\label{utu}
{\frac{1}{ 2\kappa_{11}^2}\int_{{\bf R}^4\times X} d^{11}x
\sqrt g R.}\end{equation}
 Denoting the volume of $X$ as $V_X$, this reduces in
four dimensions simply to 
\begin{equation}\label{nutu}
{\frac{V_X}{
2\kappa_{11}^2}\int_{{\bf R}^4} d^4x \sqrt g R.}\end{equation}
 The
four-dimensional Einstein-Hilbert action is 
\begin{equation}\label{gutu}
{\frac{1}{
16\pi G_N}\int_{{\bf R}^4}d^4x \sqrt g R.}\end{equation}
So
\begin{equation}\label{mutu}
{G_N=\frac{\kappa_{11}^2}{ 8\pi V_X}.}\end{equation}

Now let us work out the correctly normalized Yang-Mills action on
${\bf R}^4\times Q$. For a system of $n$ Type IIA D6-branes, the
Yang-Mills action (see eqns. (13.3.25) and (13.3.26) of
\cite{polchinski}) is 
\begin{equation}\label{ploko}
{\frac{1}{
4(2\pi)^4g_s(\alpha')^{3/2}}\int d^7x \sqrt g \Tr
F_{\mu\nu}F^{\mu\nu},}\end{equation}
 where $g_s$ is the string coupling constant
and  $\Tr $ is the trace in the fundamental representation of
$U(n)$. In the GUT literature, one usually writes
$F_{\mu\nu}=\sum_a F_{\mu\nu}^a Q_a$, where $Q_a$ are generators
of $U(n)$ normalized to $\Tr \,Q_aQ_b=\half\delta^{ab}.$  So
(\ref{ploko})\ can be written 
\begin{equation}\label{noko}
{ \frac{1}{
8(2\pi)^4g_s(\alpha')^{3/2}}\int d^7x \sqrt g\sum_a F_{\mu\nu }^a
F^{\mu\nu\,a}.}\end{equation}
 Going to $M$-theory, the relation between
$\kappa_{11}$, $g_s$, and $\alpha'$ is (eqn. (14.4.5) of
\cite{polchinski}) 
\begin{equation}\label{ygo}
{\kappa_{11}^2=\frac{1}{
2}(2\pi)^8g_s^3(\alpha')^{9/2}.}\end{equation}
 Combining these, we see that the
Yang-Mills action in seven dimensions is 
\begin{equation}\label{gloko}
{\frac{1}{
4g_7^2} \int d^7x \sqrt g \sum_a\,F_{\mu\nu }^a F^{\mu\nu\,a} =
\frac{1}{ 8 (2\pi)^{4/3}2^{1/3}\kappa_{11}^{2/3} } \int d^7x \sqrt g
\sum_a\,F_{\mu\nu }^a F^{\mu\nu\,a}.}\end{equation}
  The conventional action in
four dimensions is 
\begin{equation}\label{ugloko}
{\frac{1}{ 4 g_{GUT}^2}\int
d^4x\, \sqrt{g}\, F_{\mu\nu }^a F^{\mu\nu\,a},}\end{equation}
 with
$\alpha_{GUT}=g_{GUT}^2/4\pi$.  So after reducing (\ref{gloko})\ to four
dimensions on ${\bf R}^4\times Q$ and getting a factor of $V_Q$,
the volume of $Q$, from the integral over $Q$,  we identify
\begin{equation}\label{alphag}
{\alpha_{GUT}=\frac{(4\pi)^{1/3}\kappa_{11}^{2/3}}{
V_Q}.}\end{equation}

Combining (\ref{mutu})\ and (\ref{alphag}), we have
\begin{equation}\label{hugu}
{G_N=\frac{\alpha_{GUT}^3 V_Q^{2/3}}{ 32\pi^2 a},}\end{equation}
 with
$a=V_X/V_Q^{7/3}$. Using (\ref{timely}), we can write this as 
\begin{equation}\label{kugul}
{
G_N=\frac{\alpha_{GUT}^3 L(Q)^{2/3}}{ 32 \pi^2 a M_{GUT}^2}.}\end{equation}
 Here
$M_{GUT}$ is the unification scale as inferred from low energy
data (but if there are extra light particles not presently known,
they must be included in the extrapolation).  For the simplest
lens space, $L(Q)=4q\sin^2(5\pi w/q)$, as noted in (\ref{gudu}). As we
explained in discussing eqn. (\ref{plon}), we consider it plausible that
for many phenomenologically interesting manifolds of $G_2$
holonomy, there is an upper bound on $a$ that is of order 1.  If
this is so (and of course it could be proved in principle, and
the upper bound on $a$ computed, for a given $G_2$ manifold $X$),
then (\ref{kugul})\ is a lower bound on $G_N$ that depends on the values
of the GUT parameters, the readily computed constant $L(Q)$, as
well as the more problematic bound on $a$.

Within the uncertainties, such a bound may well be saturated in
nature.  If we use the often-quoted values $M_{GUT}=2.2\times
10^{16}$ GeV and $\alpha_{GUT}\sim 1/25$, then, with
$G_N=6.7\times 10^{-39}\,{\rm GeV}^{-2}$, we need approximately
$L(Q)^{2/3}/a=15$.  We recall, however, that these values
correspond to a minimal three family plus Higgs boson spectrum
below the GUT scale, and that the doublet-triplet splitting
mechanism \cite{gwitten}\ that we are assuming in the present paper
really leads to extra light fields in vector-like $SU(5)$
multiplets.  If we use the values $M_{GUT}\sim 8\times 10^{16}$
GeV, $\alpha_{GUT}\sim .2$, which are typical values found  in
\cite{babupati}\ for certain models that have TeV-scale vector-like
fields transforming as $\bar {\bf 5}\oplus {\bf 10}$ plus their
conjugates, then we get $L(Q)^{2/3}/a=1.7$.  Raising the value of
$\alpha_{GUT}$ to $.2$ or $.3$ will also significantly reduce the
membrane action and so alleviate the problems with quark and
lepton masses that we consider in section 6.

Alternatively, (\ref{mutu})\ and (\ref{hugu})\ can be combined to give
\begin{equation}\label{tryo}
{\kappa_{11}^2=\frac{\alpha_{GUT}^3L(Q)^3}{ 4\pi
 M_{GUT}^9}.}\end{equation}

This formula is attractive because it is expresses the
fundamental eleven-dimensional coupling $\kappa_{11}$ in terms of
quantities -- $\alpha_{GUT}$ and $M_{GUT}$ -- about whose values
we have at least some idea from experiment, and another quantity
-- $L(Q)$ -- that is readily calculable in a given model.

The eleven-dimensional Planck mass $M_{11}$ has been defined
(\cite{polchinski}, p. 199) by $2\kappa_{11}^2 = (2\pi)^8 M_{11}^{-9}$.
So we can express (\ref{tryo})\ as a formula for $M_{11}$:
\begin{equation}\label{incoc}
{M_{11}=\frac{2\pi M_{GUT}}{ \alpha_{GUT}^{1/3}
L(Q)^{1/3}}.}\end{equation}
 One important result here is that $M_{GUT}$ is
parametrically smaller than $M_{11}$ -- by a factor of
$\alpha_{GUT}^{1/3}$.  This factor of $\alpha_{GUT}^{1/3}$ is the
reason that it makes sense to use perturbation theory  -- as we
have done in computing threshold corrections in section 3.\footnote{It
also means that the radius of $Q$ and presumably of $X$ is of
order $\alpha_{GUT}^{-1/3}$ in eleven-dimensional Planck units, so
that membrane instanton actions (which scale as length cubed) are
of order $1/\alpha_{GUT}$. This fact will be troublesome in
section 6.}

Regrettably, the precise factors in the definition of $M_{11}$
have been chosen for convenience.  We really do not know if the
characteristic mass scale at which eleven-dimensional supergravity
breaks down and quantum effects become large is $M_{11}$, or $2\pi
M_{11}$, or for that matter $M_{11}/2\pi$. 
\footnote{Just to get a feel for what the characteristic mass scale might be, note that for $(q,w)=(2,1)$, we have  $M_{11}=2.0\times 10^{17}${\rm GeV} for the often-quoted values of $M_{GUT}$ and $\alpha _{GUT}$, and $M_{11}=4.3\times 10^{17}${\rm GeV} for the values in \cite{babupati}\ .}
 This uncertainty will
unfortunately be important in section 5.

\section{Proton Decay}
\label{sec:5}

In this section, we will analyze the gauge boson contribution to
proton decay in the present class of models.

First, we recall how the analysis goes in four-dimensional GUT's.
We will express the analysis in a way that is convenient for the
generalization to ${\bf R}^4 \times Q$. The gauge boson
contribution to proton decay comes from the matrix element of an
operator product 
\begin{equation}\label{dorry}
{g_{GUT}^2\int d^4x\, J^\mu(x)  \tilde
J_\mu(0)D(x,0),}\end{equation}
 where
 $J$ and $\tilde J$ are the
currents in emission and absorption of the color triplet gauge
bosons. We have used translation invariance to place one current
at the origin, and $D(x,0)$ is the  propagator of the heavy gauge
bosons that transform as $(\3,\2)^{-5/6}$ of $SU(3)\times
SU(2)\times U(1)$. Because the proton is so large compared to the
range of $x$ that contributes appreciably in the integral, we can
replace $J^\mu(x)$ by $J^\mu(0)$, and then use 
\begin{equation}\label{torry}
{\int
d^4x\, D(x,0)=\frac{1}{ M^2}}\end{equation}
 (with $M$ the mass of the heavy gauge
bosons) to reduce (\ref{dorry})\ to 
\begin{equation}\label{forry}
{\frac{g_{GUT}^2J_\mu\tilde
J^\mu(0) }{ M^2}.}\end{equation}
 (\ref{torry})\ is a direct consequence of the
equation for the propagator, which is 
\begin{equation}\label{polop}
{\left(\Delta
+M^2\right)D(x,0)=\delta^4(x),}\end{equation}
 with
$\Delta=-\eta^{\mu\nu}\partial_\mu\partial_\nu$ the Laplacian. Of
course, in deriving (\ref{torry}), we should be careful in defining the
operator product $J_\mu \tilde J^\mu$; doing so leads to some
renormalization group corrections to the above tree level
derivation. These can be treated the same way in four dimensions
and in the $G_2$-based models, and hence need not concern us here.

In ${\bf R}^4\times Q$, the idea is similar, except that the
currents are localized at specific points on $Q$, which we will
call $P_1$ and $P_2$.  The gauge boson propagator is a function
$D(x,P;y,P')$ with $x,y\in {\bf R}^4$, and $P,P'\in Q$; the
equation it obeys is 
\begin{equation}\label{gugu}
{ \left(\Delta_{{\bf
R}^4}+\Delta_Q\right)D(x,P;y,P')=\delta^4(x-y)\delta(P,P').}\end{equation}
 Here
$\Delta_{{\bf R}^4}$ is the Laplacian on ${\bf R}^4$, acting on
the $x$ variable, and similarly $\Delta_Q$ is the Laplacian on
$Q$, acting on $P$.  Now we set $P$ and $P'$ to be two of the
special points $P_1$ and $P_2$ on $Q$ (with enhanced singularities
in the normal directions) at which chiral matter fields are
supported. The analog of (\ref{dorry})\ is 
\begin{equation}\label{pugu}
{g_7^2\int d^4x
J_\mu(x,P_1)\tilde J^\mu(0,P_2)D(x,P_1;0,P_2),}\end{equation}
 where we have used
translation invariance to set $y=0$.  Again, because the proton is
so large compared to the range of $x$ that contributes
significantly to the integral, we can set $x$ to 0 in
$J_\mu(x,P_1)$, giving us 
\begin{equation}\label{nugu}
{g_7^2J_\mu(0;P_1) \tilde
J^\mu(0;P_2)\int d^4x \,D(x,P_1;0,P_2).}\end{equation}
 Now it follows from
(\ref{gugu})\ that the function 
\begin{equation}\label{turry}
{F(P,P')=\int d^4x
\,D(x,P;0,P')}\end{equation}
 obeys 
\begin{equation}\label{curryco}
{\Delta_Q F(P,P')=\delta(P,P').}\end{equation}
In other words, $F$ is the Green's function of the scalar
Laplacian on $Q$ (for scalar fields valued in the $(\3,\2)^{-5/6}$
representation).  In particular, $F$ is bounded for $P$ away from
$P'$, and for $P\to P'$, 
\begin{equation}\label{urryco}
{F(P,P')\to \frac{1}{ 4\pi
|P-P'|},}\end{equation}
 with $|P-P'|$ denoting the distance between these two
points. The proton decay interaction  is 
\begin{equation}\label{ohugu}
{g_7^2
J_\mu(0;P_1)\tilde J^\mu(0;P_2)\, F(P_1,P_2).}\end{equation}
 More exactly, this
is the contribution for fermions living at the points $P_1$,
$P_2$. It must be summed over possible $P_i$.

Given (\ref{urryco}), if it is possible to have $P_1$ very close to
$P_2$, this will give the dominant contribution.   But how close
will the $P_i$ be? The smallest that the denominator in (\ref{urryco})\
will get is if $P_1=P_2$, in other words if the currents $J_\mu$
and $\tilde J^\mu$ in the proton decay process act on the same
${\bf 10}$ or $\bar {\bf 5}$ of $SU(5)$, living at some point
$P=P_1=P_2$ on $Q$.  If this is the case, then the result (\ref{nugu})\
is infinite. $M$-theory will cut off this infinity, but we do not
know exactly how. The best we can say is that the cutoff will
occur at a distance of order the eleven-dimensional Planck length,
as this is the only scale that is relevant in studying the conical
singularity at $P$. If we naively say that setting $P_1=P_2$ means
replacing $1/|P_1-P_2|$ by $1/R_{11}=M_{11}$ (with $R_{11}$ the
eleven-dimensional Planck length; $M_{11}$ was evaluated in
(\ref{incoc})), then we would replace $F(P,P)$ by $M_{11}/4\pi$.
Unfortunately, as noted at the end of section 4, we have no idea
whether $M_{11}$ or some multiple of it is the natural cutoff in
$M$-theory. This is an important uncertainty, since (for example)
$4\pi$ is a relatively large number and the proton decay rate is
proportional to the square of the amplitude! All we can say is
that the effective value of $F(P,P)$, though uncalculable with
the present understanding of $M$-theory, is model-independent; it
does not depend on the details of $X$ or $Q$, but is a universal
property of $M$-theory with the conical singularity $P$. The
effective interaction is 
\begin{equation}\label{plook}
{\sum_{P}C\frac{M_{11}g_7^2}{
4\pi}J_\mu\tilde J^\mu(P),}\end{equation}
 where $C$ is a constant that in
principle depends only on $M$-theory and not the specific model.
So $C$  might conceivably be computed in the future (if better
methods are discovered) without knowing how to pick the right
model.  We have made explicit the fact that the interaction is
summed over all possibilities for $P=P_1=P_2$. Subleading (and
model-dependent) contributions with $P_1\not= P_2$ have not been
written.

Let us compare this to the situation in four-dimensional GUT's.
The currents $J$ and $\tilde J$ receive contributions from
particles in the ${\bf 10}$ and $\bar {\bf 5}$ of $SU(5)$.  Thus
we can expand 
\begin{equation}\label{ngi}
{J_\mu\tilde J^\mu=J_\mu^{{\bf 10}}\tilde
J^{\mu \,{\bf 10}}+J_\mu^{\bar{\bf 5}}\tilde J^{\mu \,\bar{\bf
5}}+J_\mu^{{\bf 10}}\tilde J^{\mu \,\bar{\bf 5}}+J_\mu^{\bar{\bf
5}}\tilde J^{\mu \,{\bf 10}}.}\end{equation}
 Among these terms, the ${\bf
10}\cdot {\bf 10}$ operator product contributes to $p\to \pi^0
e^+_L$, $\bar{\bf 5}\cdot \bar{\bf 5}$ does not contribute to
proton decay, and the cross terms contribute to $p\to \pi^0e^+_R$
and $p\to\pi^+\bar\nu_R$. Assuming that the points supporting
${\bf 10}$'s are distinct from the points supporting $\bar{\bf
5}$'s, the above mechanism, in comparison to four-dimensional
GUT's, enhances the decay $p\to \pi^0e^+_L$ relative to the
others.

Using the formulas in section 4, we can evaluate the product
$g_7^2M_{11}$. Reading off $g_7^2$ from (\ref{gloko})\ and $M_{11}$ from
(\ref{incoc}), we find that the effective interaction is 
\begin{equation}\label{infof}
{
\sum_{P_i} CJ_\mu\tilde J^\mu(P_i)\cdot \frac{2\pi
L(Q)^{2/3}\alpha_{GUT}^{2/3} }{ M_{GUT}^2}.}\end{equation}
 The equivalent
formula in four-dimensional GUT's is 
\begin{equation}\label{tinfof}
{J^T_\mu\tilde
J^{T\,\mu}\frac{g_{GUT}^2}{ M^2}=J^T_\mu\tilde
J^{T\,\mu}\frac{4\pi\alpha_{GUT}}{ M^2},}\end{equation}
 where the superscript $T$
refers to the total current for all fermion multiplets. Moreover,
$M$ is the mass of the color triplet gauge bosons, and so may not
coincide with $M_{GUT}$, which is the unification scale as deduced
from the low energy gauge couplings; in some simple
four-dimensional models, the ratio $M/M_{GUT}$ is computable. The
above formulas show that, in principle, the decay amplitude for
$p\to \pi^0e^+_L$ in the $G_2$-based theory is enhanced as
$\alpha_{GUT}\to 0$ by a factor of $\alpha_{GUT}^{-1/3}$ relative
to the corresponding GUT amplitude. The enhancement means that, in
some sense, in the models considered here, proton decay is not
purely a gauge theory phenomenon but a reflection of $M$-theory.

In practice, in nature, $\alpha_{GUT}^{-1/3}$ is not such a big
number.  Whether the effect we have described is really an
enhancement of $p\to \pi^0e^+_L$ or a suppression of the other
decays depends largely on the unknown $M$-theory constant $C$. The
factor $L(Q)^{2/3}$ can also be significant numerically.  For
example, for the simplest lens space, with the minimal choice
$w=1$, $q=2$, we get $L(Q)=4q\sin^2(5\pi w/q)=8$, whence
$L(Q)^{2/3}=4$.

Even if $C$ and the other factors in (\ref{infof})\ were all known, the
proton lifetime would also depend on how the light quarks and
leptons are distributed among the different $P_i$, or
equivalently, how they are distributed among the different ${\bf
10}$'s of $SU(5)$.  The same remark applies in four-dimensional
GUT's: the proton decay rate can be reduced by mixing of quarks
and leptons among themselves as well as with other multiplets,
including multiplets that have GUT-scale masses.

The arrangement of the different quarks and leptons among the
$P_i$ will also affect the flavor structure of proton decay.  What
we have referred to as $p\to \pi^0 e^+_L$ will also contain an
admixture of other modes with $\pi^0$ replaced by $K^0$, and/or
$e^+_L$ replaced by $\mu^+_L$.  As in four-dimensional GUT's,
these relative decay rates are model-dependent.

\section{Difficulties With The Model}
\label{sec:6}

Finally, in this concluding section, we will explain some of the
difficulties in making a realistic model of physics based on the
class of models that we have been exploring.

One key cluster of issues is common to all known string and
$M$-theory approaches to phenomenology.  These center around the
need for a good mechanism of  supersymmetry breaking that solves
problems such as the SUSY flavor problem, the smallness of the
cosmological constant, and the high degree of stability of the
vacuum we live in. We will not say anything here about these
general issues, and instead focus on issues that are more or less
special to the particular $G_2$ framework.

In the models described in \cite{gwitten}\ and further explored here,
$SU(5)$ multiplets $M_i$ containing Higgs bosons, quarks and
leptons, and possibly messengers of gauge-mediated supersymmetry
breaking are supported at points $P_i$ on $Q$.  As long as the
$P_i$ are distinct, superpotential interactions that ultimately
lead to quark and lepton masses come from membrane instanton
corrections and hence are exponentially small as $\alpha_{GUT}\to
0$.  (For a study of superpotentials from membrane instantons on
smooth $G_2$ manifolds, see \cite{hmt}.)

In fact, the membrane action scales as $({\rm length})^3$, and
hence, since we saw in  section four that $R_Q\sim
{\alpha_{GUT}}^{-1/3}$, the membrane action scales as
$1/\alpha_{GUT}$ if all lengths in $X$ are scaled the same way. We
would need to know details about $X$ in order to compute the
coefficients of $1/\alpha_{GUT}$ for various membrane instantons,
so it is hard to be specific here. On the plus side of the ledger,
small changes in membrane instanton actions could produce a wide
range of quark and lepton masses, such as is seen in nature.

Quarks and leptons of the first two generations -- and especially
the first -- have very small masses, but probably not as small as
one might get from a membrane instanton with an action of order
$1/\alpha_{GUT}$ if we take the usual value $\alpha_{GUT}\sim
1/25.$  So it is probably necessary to assume that, because of
extra $SU(5)$ multiplets that survive below the GUT scale,
$\alpha_{GUT}$ is significantly larger than the usual estimate.
For a proposal with extra light vector-like matter leading to
$\alpha_{GUT}\sim .2-.3$, see \cite{babupati}\ .
(We
recall that at least some extra light vector-like matter --
possibly serving as messengers of gauge-mediated supersymmetry
breaking -- is needed in the doublet-triplet splitting framework
of \cite{gwitten}\ because of certain anomalies. See also \cite{GhilenceaMU} .)

The top quark mass is so large that it is not plausible to
interpret it as being subject to any exponential suppression at
all. So we might want to assume that the point $P_1$ that
supports a Higgs ${\bf 5}$ of $SU(5)$ coincides with the point
$P_2$ that supports one of the ${\bf 10}$'s.  In a case with such
multiple singularities, one obtains a superpotential coupling
(hopefully ${\bf 5}\cdot{\bf 10}\cdot{\bf 10}$ in this example)
that is ``of order one,'' with no exponential suppression at all.
(This is shown in an example in \cite{pwitten}.) This might give us a
large top quark mass. However, if we want to split Higgs doublets
and triplets using the mechanism described in \cite{gwitten}, we really
must take the Higgs $\overline{\bf 5}$ to be supported at a point
$P_3$ that is distinct from $P_1$ and $P_2$ (in fact, $P_3$ and
$P_1$ lie on  different components of the fixed point set of the
global symmetry that leads to doublet-triplet splitting). So the
bottom quark mass presumably has at least some exponential
suppression involving a membrane instanton action.  Since
$m_b/m_t\sim 1/40$, which is not all that small, and $m_b$ can be
suppressed  by large $\tan\beta$ in supersymmetric models as well
as by a large membrane instanton action, in practice we will have
to suppose that $\tan\beta$ is not too large and than a certain
membrane instanton has rather small action.

Light neutrino masses come from dimension five superpotential
couplings (of the form $\int d^2\theta H^2 L^2$, with $H$ and $L$
being Higgs boson and lepton doublets), which again arise from
membrane instantons, or possibly by integrating out a heavy
singlet (``right-handed neutrino'').  So it is necessary to
arrange that the membrane instanton generating the $H^2L^2$
coupling has rather small action, or that the heavy singlet is
sufficiently light and sufficiently strongly coupled to $H$ and
$L$.  (The singlet mass and couplings to $HL$ both come from
membrane instantons.)

Since the $SU(5)$ relation $m_b= m_\tau$, which is converted by
renormalization group running to the weak scale  to something
more like $m_b=3m_\tau$ \cite{nano}, is fairly successful, we presumably want
to preserve this relation by avoiding significant mixing of the
third generation with other multiplets. Since analogous relations
for the first two genreations are not successful, we probably do
want mixing of the first two generations with other multiplets
that have large masses (the use of such mixing to modify fermion
mass relations was recalled in \cite{gwitten}).  It is consistent to
have significant mixing of the first two generations with other
multiplets while avoiding such mixing for the third generation,
because the observed values of the CKM quark mixing angles
suggest that (except for neutrinos) the mixing of the third
generation with the first two is very tiny. We do not have a good
mechanism that would split off the third generation in this way.

Finally, we come to the question of whether the Higgs triplet mass
$m_T$ can really be as light as suggested in (\ref{guny}).  In
four-dimensional GUT's this might lead to trouble because of
proton decay interactions mediated by the Higgs triplets.  In the
present context, the doublet-triplet splitting mechanism of
\cite{gwitten}\ implies that, because of an exotic discrete symmetry,
the Higgs triplet exchange does not generate dimension five
operators. We also have to worry about dimension six operators
mediated by the Higgs triplets; these are unavoidable, and give
proton decay amplitudes proportional to
$\lambda_1\lambda_2/m_T^2$, where the $\lambda_i$ are Yukawa
couplings of the Higgs triplet to quarks and leptons. The
$\lambda_i$ would plausibly\footnote{This estimate for the Yukawa
couplings is certainly valid if we ignore mixing of the Higgs
bosons and the first two generations with other multiplets.} be
of order $m_{q,l}/m_W$, where $m_{q,l}$ is a first or second
generation quark and lepton mass and $m_W$ is the $W$ boson
mass.  Since the ratios $m_{q,l}/m_W$ range from roughly
$10^{-4}$ to $10^{-2}$, with values closer to $10^{-4}$ for the
quarks that are part of the initial state in a proton decay
process, this is plausibly enough suppression so that we can
accept a Higgs triplet mass as low as $10^{-2}$ $M_{GUT}$. In
fact, such a Higgs triplet might not dominate proton decay,
leaving the mechanism explored in section 5 as the dominant
mechanism.

{\it Acknowledgements}

We thank M. Goresky, I. R. Klebanov, L. Randall, and P. Sarnak for
helpful discussions.

The work of T.F. is supported in part by a National Science
Foundation Graduate Research Fellowship, in part by a Soros
Fellowship for New Americans, and in part by the National Science
Foundation under Grant PHY-9802484. T.F. also thanks the
organizers of the Cargese Summer School 2002, where some of this
work was carried out. The work of E.W. is supported in part by NSF
Grant PHY-0070928.

Any opinions, findings, and conclusions or recommendations
expressed in this material are those of the authors and do not
necessarily reflect the views of the National Science Foundation
or the Soros Foundation.

\def\KK{{\cal K}}
\def\s{{\bf S}}
\def\S{{\bf S}}
\def\Z{{\bf Z}}
\def\Lens{{\bf S}^3/{\bf Z}_q}
\setcounter{section}{0}
\renewcommand{\thesection}{\Alph{section}}

\section{ On The Ray-Singer Torsion In The Trivial Representation}

In this appendix, we will calculate $\KK_{\cal O}(\S^3)$ for the
trivial representation ${\cal O}$  directly by summing over the
non-zero eigenvalues of the Laplacian and using zeta function
regularization.  We did not find this calculation in the
literature, and wanted to fill this gap; 
for more examples of the use of zeta function regularization in studying Ray-Singer torsions, see \cite{nash} .

Then we will calculate $\KK_{\cal O}$ for the lens space
$\S^3/\Z_q$  by using the relation 
\begin{equation}\label{reln}
{\KK_{\cal
O}(\S^3)=\KK_{\cal O} (\Lens)+\sum_{\omega\not={\cal
O}}\KK_{\omega} (\Lens)}\end{equation}
where the sum is
over non-trivial representations of $\Z_q$.  For such non-trivial
representations, $\K_\omega(\Lens)$ has been computed by
evaluating the zeta function \cite{rs}.  The relation (\ref{reln})\ holds
simply because
 each eigenform on $\s^3$ is
in some representation of the $\Z _q$ action on $\s^3$, so we can
separate the sum over  eigenforms on $\s ^3$ into a sum over the
eigenforms living in the trivial representation of $\Z _q$ and
those living in the non-trivial representations.

We recall the definition\footnote{In discussing $\K(\S^3)$, we omit
the subscript ${\cal O}$ as the fundamental group of $\S^3$ is
trivial.} of $\K$: 
\begin{equation}\label{recdef}
{ \KK(\s ^3 )=\frac{3}{ 2}\log \det
\Delta_0 (\s^3) - \frac{1}{ 2}\log \det \Delta _1 (\s^3).}\end{equation}
 So we
need the eigenvalues of the Laplacian and their multiplicities for
0--forms and 1--forms on $\s ^3$. We shall calculate everything
first for a sphere of radius 1, and include the dependence on the
radius later.

The 0--forms have eigenvalues $\lambda _{0,n}=n(n+2)$ with
multiplicities $y_{0,n}=(n+1)^2$. There are two types of 1--forms:
closed ones, which have the same eigenvalues and multiplicities as
the 0--forms, and co-closed 1--forms, which have eigenvalues
$\lambda _{1,n}=(n+1)^2$ and multiplicities $y_{1,n}=2n(n+2)$ (see
(3.19) of \cite{nash}
).

The logarithm of the determinant of the Laplacian is defined by
analytic continuation using zeta functions and comparing to the
known analytic continuation of the Riemann zeta function. We begin
by writing the zeta function of the Laplacian on 0--forms or
closed 1--forms, and the zeta function of the Laplacian on
co-closed 1--forms: 
\begin{eqnarray}\label{zetazero}
\zeta_0(s)&=& \sum \frac{y_{0,n}}{
\lambda _{0,n} ^s} = \sum _{n=1}^\infty \frac{(n+1)^2}{
(n(n+2))^s},\\ \label{zetaone}
\zeta _1  (s)& =& \sum \frac{y_{1,n}}{
\lambda _{1,n} ^s}= \sum _{n=1}^\infty \frac{2n(n+2)}{ (n+1)^{2s}}.\end{eqnarray}
These converge for sufficiently large ${\rm Re}(s)$.  We want to
analytically continue them to $s=0$, after which we define
\begin{equation}\label{hdefine}
{-\zeta_0'(0) =\log \det \Delta_ 0' = \log \det \Delta
_1^{ closed} ,~~~ -\zeta _1 '(0)= \log \det \Delta _1^{
co-closed}.}\end{equation}
In terms of
these zeta functions, we have 
\begin{equation}\label{interms}
{  \frac{1}{ 2}\Bigl[
3\,\log \det \Delta_ 0' -(\log \det \Delta _1^{' \, closed}+ \log
\det \Delta _1^{' \, co-closed}) \Bigr ] = \frac{1}{ 2}\zeta _1
'(0)-\zeta_0'(0).}\end{equation}
  We wish to rewrite the sums $\zeta _0(s)$ and
$\zeta _1(s)$ in terms of the well known Riemann zeta function
$\zeta (s)$ so that they can be continued  to $s=0$. This is
straightforward for $\zeta _1(s)$: 
\begin{equation}\label{strz}
{ \zeta _1(s)= 2\sum
_{n=1}^\infty \Bigl ( \frac{(n+1)^2}{ (n+1)^{2s}}-\frac{1}{
(n+1)^{2s}}\Bigr ).}\end{equation}
 So we can write $\zeta_1(s)$ in terms of the
Riemann zeta function $\zeta(s)$: 
\begin{equation}\label{uncv}
{ \zeta _1(s)=2((\zeta
(2s-2)-1)-(\zeta (2s)-1))=2(\zeta (2s-2)-\zeta (2s)).}\end{equation}
 Since the
analytic continuation of $\zeta(s)$ is well-known, this solves
the problem of analytically continuing $\zeta_1(s)$.
 As for
$\zeta_0(s)$, we rewrite it as follows: 
\begin{equation}\label{rewriter}
\begin{aligned}
\zeta _0(s)=& \sum _{n=1}^\infty (n+1)^2
 \Bigl [\left ( 1-\frac{1}{ (2s+1)}\right )\frac{1}{ (n+1)^{2s}}
 +\frac{1}{ 2(2s+1)}\Bigl ( \frac{1}{ n^{2s}}+ \frac{1}{ (n+2)^{2s}}\Bigr )\Bigr ]
 \cr
&+ \sum _{n=1}^\infty (n+1)^2 \Bigl [ \frac{1}{ (n(n+2))^s} - \left
( 1-\frac{1}{ (2s+1)}\right )\frac{1}{ (n+1)^{2s}}\cr &\hspace{2em}-\frac{1}{
2(2s+1)}\Bigl ( \frac{1}{ n^{2s}}+ \frac{1}{ (n+2)^{2s}}\Bigr )\Bigr
] . \cr\end{aligned}\end{equation}
The sum on the second line converges absolutely for
${\rm Re} (s)>-1/2$: rewriting it with $u=\frac{1}{ n+1}$, it
becomes 
\begin{equation}\label{humdo}
{\sum _{n=1}^\infty u^{2s-2}\Bigl [ \frac{1}{
(1-u^2)^s}-\left ( 1-\frac{1}{ (2s+1)}\right )-\frac{1}{
2(2s+1)}\Bigl ( \frac{1}{ (1-u)^{2s}}+ \frac{1}{ (1+u)^{2s}}\Bigr )
\Bigr ],}\end{equation}
 and one can see that the leading order term for large
$n$ (small $u$) is $u^{2s-2}u^4=u^{2s+2}$.  Hence it is bounded by
$1/(n+1)^{2s+2}$ which converges absolutely for ${\rm Re}
(2s+2)>1$ or ${\rm Re} (s)>-1/2$. So for ${\rm Re} (s)>-1/2$, we
can do the sum term by term.  At $s=0$, each term in the sum
vanishes, and furthermore the derivative of each term with
respect to $s$ at $s=0$ vanishes. Therefore, for small $s$ we can
write 
\begin{equation}\label{gugo}
{ \zeta_0(s)= \sum _{n=1}^\infty (n+1)^2\Bigl
[\left ( 1-\frac{1}{ (2s+1)}\right )\frac{1}{ (n+1)^{2s}}+\frac{1}{
2(2s+1)}\Bigl ( \frac{1}{ n^{2s}}+ \frac{1}{ (n+2)^{2s}}\Bigr ) \Bigr
],}\end{equation}
 which becomes, by analytic continuation, 
\begin{equation}\label{guxu}
{ \zeta_0(s)
=\zeta(2s-2)+\frac{1}{ (2s+1)}\zeta (2s)-\left (1-\frac{1}{
(2s+1)}\right )-\frac{1}{ (2s+1)2^{2s+1}} .}\end{equation}
 Using known values of
the Riemann zeta function 
\begin{equation}\label{jdf}
{ \zeta (-2)=0\, , \hskip 1cm
\zeta (0)=-1/2\, , \hskip 1cm \zeta ' (0)=-\frac{1}{ 2}\log 2\pi,}\end{equation}
we have at $s=0$ the following values of the zeta functions of
our Laplacians and their derivatives: 
\begin{equation}\label{torney}
\begin{aligned}
\zeta_0(0)=& \zeta (-2)+\zeta (0)-\frac{1}{ 2}= -1 ,\cr \zeta
_1(0)=& 2(\zeta (-2)-\zeta (0))=1,\cr
 \zeta '_0(0)
=&2\zeta '(-2)-\log \pi,\cr
 \zeta _1'(0)=&4(\zeta '(-2)-\zeta
'(0))=4\zeta '(-2)+2\log (2\pi). \cr\end{aligned}\end{equation}
 (The value of $\zeta
_0'(0)$ can also be obtained from Proposition 3.1 of \cite{ng}, or from
\cite{bg}.)

Now we include the dependence on the radius. The quantity that
comes into the physical calculation is actually 
\begin{equation}\label{gugu}
{ -\Tr
\log \Big ( \frac{m_{k,n}^2 }{ \Lambda ^2 }\Big ) =  -\sum _n y_n
\log \frac{\lambda _{k,n} }{ \Lambda ^2 R^2 }~,}\end{equation}
 where $\Lambda$ is
the cutoff and $R$ is the radius of $\S^3$.
 So we should replace
$\zeta_0$ and $\zeta_1$ as defined in equations (\ref{zetazero})\ and
(\ref{zetaone})\  by 
\begin{equation}\label{repleq}
\begin{aligned}\eta _0(s)=&(R\Lambda)^{2s}
\zeta _0(s) ,\cr  \eta _1 (s)=&(R\Lambda)^{2s}\zeta _1(s) .\cr\end{aligned}\end{equation}
The derivatives at $s=0$ are 
\begin{equation}\label{theder}
\begin{aligned}\eta _0 '(0) =&
-\log R^2\Lambda ^2 +\zeta _0 ' (0), \cr \eta _1'(0)=& \log
R^2\Lambda^2 +\zeta _1'(0) .\cr\end{aligned}\end{equation}
 We now have 
\begin{equation}\label{epleq}
\begin{aligned}\KK (\s^3, 1)=&\frac{1}{ 2}\eta '_1(0)-\eta '_0(0) \cr =&
\frac{3}{ 2}\log R^2\Lambda ^2 + \log (2\pi ^2)  .\cr\end{aligned}\end{equation}

\def\T{{\cal T}}
\def\K{{\cal K}}
Finally, we can use equation (\ref{reln})\ to extend this result to a
lens space. By evaluating zeta functions, Ray showed\footnote{We
express this result using the normalization of the torsion that is
used in \cite{rs}\ and in the present paper.  Also, we include the
integer $m$ described in the last paragraph of section 3.4, with
$jm\equiv 1$ mod $q$.} \cite{ray}\ that for non-trivial $\omega$,
\begin{equation}\label{hdf}
{{\cal T}_\omega(\Lens)=\log |\omega-1||\omega^{-j}-1|.}\end{equation}
Here $\omega$ is any non-trivial $q^{th}$ root of unity, and $j$
is prime to $q$, so 
\begin{equation}\label{ughu}
{ \sum _{\omega\not= {\cal O}}
\KK_\omega (\Lens ) =\sum _{\omega\not={\cal O}} {\cal
T}_\omega(\Lens)= \sum _{\omega\not={\cal O}} \log |\omega
-1||\omega ^{-j} -1|= 2\log q~. }\end{equation}
Therefore,
\begin{equation}\label{theref}
\begin{aligned}\KK_{\cal O} (\Lens)=& \frac{3}{ 2}\log
R^2\Lambda ^2+\log 2\pi ^2 -2\log q \cr =& \log \left (\frac{2\pi
^2 }{ q ^2}\right )+\frac{3}{ 2}\log R ^2\Lambda ^2. \cr\end{aligned}\end{equation}
 In
terms of the volume $V=2\pi ^2 R^3/q$ of the lens space, this
becomes 
\begin{equation}\label{hyto}
\begin{aligned} \KK_{\cal O} (\Lens )= & \log \Big (
\frac{2\pi ^2 }{ q^2}\Big )  +\log \frac{q}{ 2\pi ^2}V\Lambda ^3 \cr
=&\log \frac{V\Lambda ^3}{ q}.\cr\end{aligned}\end{equation}

The torsion of the lens space for the trivial representation is
$\T_{\cal O}(\Lens)=\log(1/q)$.  To compute this, we use the fact
that, as described in \cite{ray}, the lens space has a cell
decomposition in which the chain group $C_k$ is isomorphic to
$\Z$ for $k=0,\dots,3$.  The only non-trivial boundary operator is
$\partial_{2\to 1}:C_2\to C_1$, which equals multiplication by
$q$. To compute $\T_{\cal O}(\Lens)$, relative to a basis of the
integral homology, we should first remove subgroups of the chain
groups that generate the homology.  In this case, we do this by
dropping $C_0$ and $C_3$.  Then the Reidemeister torsion of the
lens space for the trivial representation is defined as an
alternating sum of logarithms of the boundary maps; in the
present case, this reduces to $-\log\partial_{2\to 1}=\log(1/q)$.
  The Reidemeister
torsion equals the Ray-Singer torsion by the conjecture of Ray and
Singer, which was later proved by Cheeger, so $\T_\00=\log(1/q)$.
In \cite{ray}, (\ref{hdf})\ was obtained by a similar computation.

So (\ref{hyto})\ implies that 
\begin{equation}\label{oko}
{\K_{\cal O}=\T_{\cal
O}+\log(V\Lambda^3)}\end{equation}
 for lens spaces. In the next appendix, we
show that this relation is actually true for all three-manifolds
with $b_1=0$.  This relation was used in section 3 in evaluating
the threshold corrections.

\section{The Volume Correction}

 Here we show the relation (\ref{oko})\
for every three-manifold $Q$.  In the derivation, we set
$\Lambda=1$, though from a physical point of view it is natural to
include $\Lambda$.

In defining  Reidemeister torsion for a representation that has
non-trivial real cohomology, one has to pick a basis for the
cohomology.  In the case of the trivial representation ${\cal O}$,
we can pick a basis of integral classes that generate the
integral cohomology mod torsion; the Reidemeister torsion is
independent of the choice of such a basis.  When we speak of
$\T_{\cal O}(Q)$, we mean the torsion defined relative to such a
basis, which we call a topological basis.

In Ray-Singer torsion, instead, the natural basis would be a
basis of zero modes $\alpha_i$ of the Laplacian that are
orthonormal in the ${\bf L}^2$ sense, that is, $\int_Q\bar
\alpha_i\wedge *\alpha_j=\delta_{ij}$.  We call such a basis an
${\bf L}^2$ basis. The theorem relating Ray-Singer and
Reidemeister torsion asserts that 
\begin{equation}\label{guty}
{\T_{\cal
O}(Q)=\K_{\cal O}(Q)+{\cal A},}\end{equation}
 where ${\cal A}$ is defined as
follows. For each $k=0,\dots,3$, we let $A^k$ be an invertible map
from a topological basis of the $k^{th}$ cohomology to an ${\bf
L}^2$ basis.  Then the quantities $|\det(A^k)|$ are independent
of the choices of bases and maps, and the general definition of
$\A$ is 
\begin{equation}\label{onson}
{\A=\sum_{k=0}^3(-1)^k\log|\det(A^k)|.}\end{equation}

For a three-manifold $Q$ of $b_1(Q)=0$, everything simplifies
drastically.  The nonzero Betti numbers are $b_0=b_3=1$.  For a
topological basis, we can pick the zero-form 1 and the three-form
$\epsilon_{ijk}dx^idx^jdx^k/V_Q$, where the volume is $\int
\epsilon_{ijk}dx^idx^jdx^k=V_Q$.  For an ${\bf L}^2$ basis, we
pick $1/\sqrt{ V_Q}$ and $\epsilon_{ijk}dx^idx^jdx^k/\sqrt{V_Q}$.
So $\det A^0=1/\sqrt{ V_Q}$, $\det A^3=\sqrt{ V_Q}$, and we arrive
at (\ref{oko}).




\end{document}